\documentclass[11pt,a4paper]{article}
\pdfoutput=1 
\usepackage{jheppub}	
\usepackage{mathabx}
\usepackage{tikz}
\usetikzlibrary{positioning,arrows}
\usetikzlibrary{decorations.pathmorphing}
\usetikzlibrary{decorations.markings}
\usetikzlibrary{snakes}

\usepackage{epsfig}
\usepackage{graphicx}
\usepackage[vcentermath,enableskew]{youngtab}

\usepackage[utf8]{inputenc}
\usepackage[english]{babel}
\usepackage{makeidx}
\usepackage{tikz}
\tikzset{every picture/.style={line width=0.75pt}} 
\tikzset{line/.style={thick, decorate, draw=black,}}
\usepackage{amsfonts}
\usepackage{enumerate}
\usepackage{mathrsfs}
\usepackage{tensor}
\usepackage[autostyle]{csquotes}
\usepackage{subfig}

\def\be{\begin{equation}}
\def\ee{\end{equation}}
\def\bea{\begin{eqnarray}}
\def\eea{\end{eqnarray}}
\newcommand{\nn}{\nonumber}

\newcommand{\ft}[2]{{\textstyle\frac{#1}{#2}}}
\def\ii{{\rm i}}

\def\apjl{\ref@jnl{ApJ}}

\numberwithin{equation}{section}

\usepackage[utf8]{inputenc}

\date{\today}

\def\be{\begin{equation}}
\def\ee{\end{equation}}

\newcommand{\diff}{\mathrm{d}}

\newmuskip\pFqmuskip

\newcommand*\pFq[6][8]{%
  \begingroup 
  \pFqmuskip=#1mu\relax
  \mathchardef\normalcomma=\mathcode`,
  \mathcode`\,=\string"8000
  \begingroup\lccode`\~=`\,
  \lowercase{\endgroup\let~}\pFqcomma
  {}_{#2}F_{#3}{\left(\genfrac..{0pt}{}{#4}{#5}\Big | #6\right)}%
  \endgroup
}
\newcommand{\pFqcomma}{{\normalcomma}\mskip\pFqmuskip}

\newcommand{\pmat}{\begin{pmatrix}}
\newcommand{\fpmat}{\end{pmatrix}}
\newcommand{\eq}{\begin{equation}}
\newcommand{\feq}{\end{equation}}
\newcommand{\cas}{\begin{cases}}
\newcommand{\fcas}{\end{cases}}

\newcommand{\eqarray}{\begin{eqnarray}}
\newcommand{\feqarray}{\end{eqnarray}}

\newcommand{\g}{\gamma}

\def\a{\alpha}				\def\g{\gamma}		\def\d{\delta}
						
		\def\k{\kappa}				
						\def\p{\pi}				
								
\def\c{\chi}

						\def\S{\Sigma}

\title{Rotating Topological Stars}

\author[]{Massimo Bianchi, Giuseppe Dibitetto, Jose F.  Morales and Alejandro Ruip\'erez}

\affiliation[]{Sezione INFN Roma Tor Vergata and Dipartimento di Fisica, Università di Roma Tor Vergata, Via della Ricerca Scientifica 1, 00133, Roma, Italy}

\abstract{We construct a three-parameter family of smooth and horizonless rotating solutions of Einstein-Maxwell theory with Chern-Simons term in five dimensions and discuss their stringy origin in terms of three-charge brane systems in Type IIB and M-theory.  The general solution interpolates smoothly between Kerr and static Topological Star geometries. We show that for specific choices of the parameters and quantized values of the angular momentum the geometry terminates on a smooth five-dimensional cap, and it displays neither ergoregion nor closed timelike curves.  We discuss the propagation of particles and waves showing that geodetic motion is integrable and the radial and angular propagation of scalar perturbations can be separated and described in terms of two ordinary differential equations of confluent Heun type.}

\emailAdd{bianchi,dibitetto,morales,ruiperez@roma2.infn.it}

\def\be{\begin{equation}}
\def\ee{\end{equation}}
\def\bea{\begin{eqnarray}}
\def\eea{\end{eqnarray}}
\begin{document}
\tikzset{
line/.style={thick, decorate, draw=black,}
 }

\maketitle

\section{Introduction}
Deciphering the intime structure of  black holes (BHs) is one of the most fascinating and active endeavours in the quest for a quantum theory of gravity. One of the most credited proposals \cite{Mathur:2005zp}, motivated by String Theory, posits the absence of both the horizon and the curvature singularity. In this framework, BHs are viewed as statistical ensembles involving a huge number of smooth and horizonless micro-state geometries, known as \emph{fuzzballs}.  Given the existence of no-go theorems in four dimensions \cite{Einstein:1943ixi}, fuzzball geometries are typically built as solutions of five or higher-dimensional gravity theories, blowing up near the would-be horizon, see \cite{Giusto:2004id, Giusto:2004ip, Bena:2005va, Bena:2007kg, Bena:2015bea, Bah:2021jno} for a few examples within this class. 

A particularly simple class of solutions, that may be viewed as a ``caricature" of spherically-symmetric fuzzballs, are known as Topological Stars \cite{Bah:2020ogh, Bah:2020pdz, Bah:2022yji, Chakraborty:2025ger} (Top Stars or TS for brevity). TS are static, smooth and horizonless solutions of five-dimensional Einstein-Maxwell theory supported by fluxes. Several dynamical properties of these solutions have been studied in the recent literature \cite{Heidmann:2022ehn, Bianchi:2023sfs, Heidmann:2023ojf, Bena:2024hoh, Melis:2024kfr, Dima:2025zot, Bianchi:2024vmi, Bianchi:2024rod, DiRusso:2025lip}, including their linear stability, deformability, echoes due to late-time emission of (scalar) waves (as a proxy for gravitational waves) and radiation losses in the Self-Force approach, both for bound and unbound orbits. 
    
Given the wealth of data that the LIGO-Virgo-KAGRA collaboration is collecting, and the one expected from the next generation of both ground-based and space-based experiments, one may envisage the possibility in the near future of being able to discriminate black holes from fuzzballs or other Exotic Compact Objects (ECO) from their gravitational-wave signals \cite{ Bianchi:2020bxa, Bena:2020see, Bianchi:2020miz, Bena:2020uup, Ikeda:2021uvc, Bah:2021jno, Staelens:2023jgr}. For this reason it is urgent to find smooth horizonless counterparts of Kerr BHs in the simple-minded (`caricature') version of the fuzzball proposal for compact rotating objects.

The aim of the present investigation is to built a class of smooth horizonless solutions of minimal supergravity in five dimensions carrying non-trivial angular momentum in four-dimensions after reduction on a compact circle.  A three parameter family of solutions of this theory parametrized by a mass, a rotation parameter $a$ and a duality phase was found in \cite{Compere:2009zh}.  Here we consider solutions obtained from this family by analytic continuation to the imaginary branch where $a^2<0$.  The resulting solution, parametrized by three real numbers, $r_s$, $r_b$ and $a^2$  interpolate between the static TS solution at $a\to 0$, and the  Kerr BH at $r_b\to 0$. We find that for specific choices of the parameters the BH curvature singularity can be hidden inside of a cap where space terminates. Moreover, for quantized choices of the angular momentum parameter, in the branch where  $a^2<0$, the resulting geometry is horizonless and smooth everywhere.  We will refer to these solution as a ``Rotating Topological Star" (RTS for brevity). 
  
 We will study the properties of RTS, their reduction to four dimensions, and their embedding in String theory as `harmonic superpositions' of three stacks of M5-branes in M-theory or, as bound-states of KK-monopoles and D1-branes and D5-branes in Type IIB superstring theory. We will also briefly discuss the propagation of scalar and gravitational waves in the RTS geometry. In particular we show that, like in the Kerr case, the scalar wave equation can be separated into two ordinary differential equations of confluent Heun type. 
  
The plan of the paper is the following. In section 2, we discuss the M- and String origin of static Top(ological) Stars (TS).  In section 3, we introduce the RTS solutions and determine the conditions under which the geometry is horizonless and smooth. We also show that no ergoregion is present in the RTS regime.   In section 4, we discuss the reduction to four dimensions and compute mass, charges and angular momentum that turn out to be incompatible with the BH regime. In section 5 we discuss geodetic motion, show that it's integrable and identify the light-rings in the equatorial plane. Finally in Section 6 we study scalar wave perturbations and show that both the angular and radial dynamics are governed by Confluent Heun Equations (CHE's) that can be related to quantum Seiberg-Witten curves for ${\cal N}=2$ super Yang Mills (SYM) theories with three fundamental hyper-multiplets. We also discuss the extremal limit $r_b=r_s$.

\section{M- and String origin of static Topological Stars}
Let us consider the Einstein-Maxwell action in five dimensions,
\begin{equation}
S\,=\,\frac{1}{2\k_5^2}\int \diff^5 x \sqrt{-g}\left(R-\frac{1}{4}|F_{(2)}|^2 \right)\,, 
\end{equation}
where $\k_5^2=8\pi G_5$ denotes the five-dimensional gravitational coupling constant and $F_{(2)}\,=\, \diff A_{(1)}$. A   solution of this theory, found by Horowitz and Strominger in \cite{Horowitz:1991cd}, is the following:
\begin{equation}
\begin{aligned}
\diff s^2 \,=\,&-  f_s(r)\diff t^2+\frac{\diff r^2}{f_s(r)f_b(r)}+r^2\left(\diff \theta^2+\sin^2\theta \diff \phi^2\right)+f_b(r) \diff y^2\ , \\[1mm]
A_{(1)}\,=\,&  P\, \cos \theta \,\diff \phi\ ,\\[1mm]
\end{aligned}
\label{TS_5d}
\end{equation}
where 
\begin{equation}
\begin{aligned}
f_s(r) \,=\, \left(1-\frac{r_s}{r}\right) \,, \hspace{1cm} f_b(r) \,= \,\left(1-\frac{r_b}{r}\right)\,,
\end{aligned}
\end{equation}
 and 
 \begin{equation}
 P^2 \,=\, 3 \,r_s r_b\, .
 \end{equation}
In the regime $r_s\ge r_b$, which was the one originally studied in \cite{Horowitz:1991cd}, the solution describes a magnetically-charged black string.  The limiting case $r_s=r_b$ represents an extremal black string, which displays an AdS$_{3}\times S^2$ near-horizon geometry. The analysis of the solution in the regime $r_b>r_s$ has been recently addressed by Bah and Heidmann in \cite{Bah:2020ogh}, where it has been shown to correspond to a smooth horizonless soliton, called `Topological Star' or Top Star (TS) for brevity. 

There exists also an electric version of the Topological Star, where the metric  (\ref{TS_5d}) is instead supported by the electric two-form charge 
\begin{equation}
\begin{aligned}
B_{(2)}\,=\,& \frac{Q}{r} \,\diff t \wedge \diff y \ ,
\end{aligned}
\label{TS_5d2}
\end{equation}
  solving the Einstein equation  describing gravity minimally coupled to the three-form field strength  $H_{(3)}=dB_2$ 
\begin{equation}
S_5\,=\,\frac{1}{2\k_5^2}\int \diff^5 x \sqrt{-g}\left(R -\frac{1}{12}|H_{(3)}|^2\right)\,, 
 \end{equation}
The solution we have just presented can be obtained from three-charge configurations consisting of non-extremal brane systems in string or M-theory, upon a suitable identification of the charges. 
These can be viewed as dyonic solutions of minimal supergravity, obtained by identifying the three U(1) gauge fields of the  STU supergravity model in five dimensions \cite{Sierra:1985ax}, thus extending in the bosonic sector Einstein-Maxwell (EM) with a Chern-Simons (CS) term.
Denoting by $Q_i$ and $P_i$, the corresponding electric and magnetic charges with respect to the three U(1) gauge fields of the STU model, Einstein equations require 
\begin{equation}
 \sum_{i=1}^3 (P_i^2+Q_i^2) =\, 3 r_s r_b\, .
 \end{equation}
 We start by reviewing the non-extremal deformations of BPS M-branes, and then we discuss in turn the embedding  of electric and magnetic TS in string/M-theory.

\subsection{Non-BPS deformations of M-branes}
  
In this section we review  two different non-extremal deformations of BPS M-branes. An analogous story applies, \emph{mutatis mutandis}, to D-branes in string theory as well. We start by considering a BPS M5-brane described by the following eleven-dimensional supergravity solution \cite{Gueven:1992hh, Ortin:2015hya},
\begin{equation}
\begin{array}{lcl}
\diff s_{11}^2 &=& H^{-1/3}\diff s_{\mathbb{R}^{1,5}}^2 + H^{2/3}\left(\diff r^2+r^2 \diff s_{S^4}^2\right) \ , \\[2mm]
A_{(6)} &=& (H^{-1}-1)\, \mathrm{vol}_{\mathbb{R}^{1,5}} \ ,
\end{array}
\end{equation}
where $H$ is a harmonic function on the transverse $\mathbb{R}^5$ and $\diff s_{S^4}^2$ is the metric of a unit $S^4$. Demanding spherical symmetry, 
\begin{equation}
H=1+\frac{Q_5}{r^3}\, .
\end{equation}
The first way to make the above M5 become non-extremal is to add a blackening factor,
\begin{equation}
W_1=1-\frac{\omega_1}{r^3}\, ,
\end{equation}
with $\omega_1>0$ through 
\begin{equation}
\begin{array}{lcl}
\diff s_{11}^2 &=& H^{-1/3}\left(-W_1\,\diff t^2+\diff s_{\mathbb{R}^5}^2\right) + H^{2/3}\left(\frac{\diff r^2}{W_1}+r^2 \diff s_{S^4}^2\right) \ , \\[2mm]
A_{(6)} &=& \g_1\, (H^{-1}-1)\, \mathrm{vol}_{\mathbb{R}^{1,5}} \ ,
\end{array}
\end{equation}
with $\g_1=\sqrt{1+\frac{\omega_1}{Q_5}}$ the off-extremality parameter. A second way of going non-extremal is to add a cap-off factor, 
\begin{equation}
W_2\,=\,1-\frac{\omega_2}{r^3}\, ,
\end{equation}
still with $\omega_2>0$. The solution is modified as follows
\begin{equation}
\begin{array}{lcl}
\diff s_{11}^2 &=& H^{-1/3}\left(W_2\,\diff y^2+\diff s_{\mathbb{R}^{1,4}}^2\right) + H^{2/3}\left(\frac{\diff r^2}{W_2}+r^2 \diff s_{S^4}^2\right) \ , \\[2mm]
A_{(6)} &=& \g_2\, (H^{-1}-1)\, \mathrm{vol}_{\mathbb{R}^{1,5}} \ ,
\end{array}
\end{equation}
where $\g_2=\sqrt{1+\frac{\omega_2}{Q_5}}$ the off-extremality parameter.  Note that in this case space terminates at $r=\omega_2^{1/3}$, provided that the $y$ coordinate be compact. It is worth noticing that, both non-extremal M5-branes are obtained by replacing the eleven-dimensional flat space where the M5 branes live in with (Euclidean) Schwarzschild-Tangherlini (ST) Ricci-flat geometries:
 $\textrm{ST}_{1,5}\times\mathbb{R}^5$, for the former, or ${\mathbb{R}^{1,4}}\times\textrm{EST}_6$, for the latter.
We will show that both of these non-extremal deformations may be performed simultaneously in three-charge brane systems, just at the price of slightly modifying the ansatz for the flux.

\subsection{M-theory embedding of a magnetic Topological Stars}
Let us consider the intersection of three mutually orthogonal M5-branes as shown in table \ref{Table:Mbranes}.
\begin{table}[h!]
\renewcommand{\arraystretch}{1}
\begin{center}
\scalebox{1}[1]{
\begin{tabular}{c||c c|c c c c c c|c c c}
branes & $t$ & $y$ & $x^1$ & $x^{2}$ & $x^{3}$ & $x^{4}$ & $x^5$ & $x^6$ & $r$ & $\theta$ & $\phi$  \\
\hline \hline
M5$_{{1}}$ & $-$ & $-$ &  $\cdot$  & $\cdot$  & $-$ & $-$ & $-$ & $-$ & $\cdot$ & $\cdot$ & $\cdot$  \\
M5$_{{2}}$ & $-$ & $-$ & $-$ & $-$ & $\cdot$  & $\cdot$  & $-$ & $-$ & $\cdot$ & $\cdot$ & $\cdot$ \\
M5$_{{3}}$ & $\underbrace{-}_{W_1}$ & $\underbrace{-}_{W_2}$ & $-$ & $-$ & $-$ & $-$ & $\cdot$  & $\cdot$ & ${\cdot}$ & $\cdot$ & $\cdot$  \\
\end{tabular}
}
\end{center}
\caption{{\it (Non-)extremal intersection of three orthogonal M5-branes.}} \label{Table:Mbranes}
\end{table}
The corresponding 11d supergravity solution reads 
\begin{equation}
\hspace{-7mm}
\begin{array}{lcl}
\diff s_{11}^2 &=& H_1^{-1/3}H_2^{-1/3}H_3^{-1/3}\left(-W_1\,\diff t^2+W_2\,\diff y^2\right) + H_1^{2/3}H_2^{2/3}H_3^{2/3}\left(\dfrac{\diff r^2}{W_1 W_2}+r^2 \diff s_{S^2}^2\right)  \\[2mm]
&& +H_1^{-1/3}H_2^{-1/3}H_3^{-1/3}\left(H_2\,\diff s_{\mathbb{T}_1^2}^2+H_3\,\diff s_{\mathbb{T}_2^2}^2+H_1\,\diff s_{\mathbb{T}_3^2}^2\right) \ , \\[2mm]
F_{(4)} &=& \left( \g_1\,Q_1  \mathrm{vol}_{\mathbb{T}_1^2} + \g_2 \,Q_2  \mathrm{vol}_{\mathbb{T}_2^2}+\g_3 \,Q_3  \mathrm{vol}_{\mathbb{T}_3^2}\right)\wedge \mathrm{vol}_{S^2}\ ,
\end{array}
\end{equation}
where 
\begin{equation}
H_i \,=\, 1+\frac{Q_i}{r} \ , \ i=1,2,3 \ , \quad \quad  W_a \,=\, 1-\frac{\omega_a}{r} \ , \ a=1,2 \ ,
\end{equation}
with $\ \g_i \,=\, \sqrt{\left(1+\dfrac{\omega_1}{Q_i}\right)\left(1+\dfrac{\omega_2}{Q_i}\right)}\ $.  

When $W_1=W_2=1$, the system is ${1}/{8}$ -- BPS, with near-horizon geometry $\textrm{AdS}_3\times S^2$ preserving eight real supercharges.

To obtain the TS, we identify $Q_1=Q_2=Q_3 \equiv \widehat{Q}$, leading to
\be
\diff s_{11}^2 \,=\, H^{-1}\left(-W_1\,\diff t^2+W_2\,\diff y^2\right) + H^{2}\left(\frac{\diff r^2}{W_1 W_2}+r^2 \diff s_{S^2}^2\right) +\diff s_{\mathbb{T}^6}^2 \ ,
\label{TSxT6}
\ee
with $H=1+\frac{\widehat{Q}}{r}=H_1=H_2=H_3$. The metric in \eqref{TSxT6} is the direct product of  the TS metric (\ref{TS_5d}) and a six-torus, upon identifying $r_s \,=\, \widehat{Q}+\omega_1$, $r_b\,=\, \widehat{Q}+\omega_2$, and  sending  $r\to r-\widehat{Q}$, that we will henceforth continue to denote by $r$. The metric is supported by the magnetic charge $P^2=3r_s r_b $. We stress that thanks to the identification of the charges $Q_a$ and thus of the harmonic functions $H_a$ the six-torus completely factorizes from the rest and has fixed volume/radii.

\subsection{IIB embedding of a dyonic Topological Stars}
The type IIB brane picture of a dyonic TS may be obtained from the previous M5 intersection by a chain of dualities. In particular, we first reduce on $S^1_{x^6}$ to obtain an NS5 -- D4 -- D4 intersection in type IIA. Subsequently, we perform three T-dualities along $x^{3,4,5}$ to get the IIB setup depicted in table \ref{Table:IIBbranes}.
\begin{table}[h!]
\renewcommand{\arraystretch}{1}
\begin{center}
\scalebox{1}[1]{
\begin{tabular}{c||c c|c c c c c|c c c}
branes & $t$ & $y$ & $x^1$ & $x^{2}$ & $x^{3}$ & $x^{4}$ & $x^5$  & $r$ & $\theta$ & $\phi$  \\
\hline \hline
KK5$_{\textrm{B}}$ & $-$ & $-$ & $-$ & $-$ & $-$ & $-$ & {\rm iso}  &  ${\cdot}$ & $\cdot$ & $\cdot$  \\
D1 & $-$ & $-$ & {{$\cdot$}}  & $\cdot$  & $\cdot$ & $\cdot$ & $\cdot$ &  $\cdot$ & $\cdot$ & $\cdot$  \\
D5 & $\underbrace{-}_{W_1}$ & $\underbrace{-}_{W_2}$ & $-$ & $-$ & $-$  & $-$  & $\cdot$ & $\cdot$ & $\cdot$ & $\cdot$ \\
\end{tabular}
}
\end{center}
\caption{{\it (Non-)extremal KK5 -- D1 -- D5 intersection.}} \label{Table:IIBbranes}
\end{table}

The corresponding IIB supergravity (string frame) solution reads
\begin{equation}
\hspace{-7mm}
\begin{array}{lcl}
\diff s_{10}^2 &=& H_1^{-1/2}H_2^{-1/2}\left(-W_1\,\diff t^2+W_2\,\diff y^2\right) + H_1^{1/2}H_2^{1/2}H_3\left(\dfrac{\diff r^2}{W_1 W_2}+r^2 \diff s_{S^2}^2\right)  \\[3mm]
&& +H_1^{1/2}H_2^{1/2}H_3^{-1}\left(dx^5+\g_3\, Q_3 \cos\theta \,\diff\phi\right)^2+H_1^{1/2}H_2^{-1/2} \diff s_{\mathbb{T}^4}\ , \\[2mm]
e^\Phi &=& H_1^{1/2}H_2^{-1/2}  \ , \\[2mm]
C_{(2)} &=& \g_1\,\left(H_1^{-1}-1\right)\diff t \wedge \diff y\ , \\[2mm]
C_{(6)} &=& \g_2\,\left(H_2^{-1}-1\right)\diff t \wedge \diff y \wedge \mathrm{vol}_{\mathbb{T}^4}\ ,
\end{array}
\end{equation}
where 
\begin{equation}
H_i \,=\, 1+\frac{Q_i}{r} \ , \ i=1,2,3 \ , \quad \quad  W_a \,=\, 1-\frac{\omega_a}{r} \ , \ a=1,2 \ ,
\end{equation}
with $\ \g_i \,=\, \sqrt{\left(1+\dfrac{\omega_1}{Q_i}\right)\left(1+\dfrac{\omega_2}{Q_i}\right)}\ $. 

The D1-branes are smeared along the directions $x^{1,2,3,4,5}$, while the D5-branes are smeared along the direction $x^5$, the isometric direction for the KK monopole. When $W_1=W_2=1$, the system is ${1}/{8}$ -- BPS  with near-horizon geometry $\textrm{AdS}_3\times S^2$ preserving eight real supercharges.

The TS metric (\ref{TS_5d}) is obtained again identifying the three charges  $Q_1=Q_2=Q_3 \equiv \widehat{Q}$, dimensionally reducing along the  ${\mathbb{T}^4}$ and $x^5$-directions,  setting $r_s=\widehat{Q}+\omega_1$, $r_b=\widehat{Q}+\omega_2$,  and sending $r\to r-\widehat{Q}$.  The metric is now supported by a magnetic charge $P^2=r_s r_b$ coming from the KK gauge field  with charge $\gamma_3 Q_3$ and the two two-forms descending from  the RR $C_{(2)}$ and $C_{(6)}$ potentials carrying a total electric charge $Q^2= 2r_s r_b$. As before, thanks to the identification of the charges $Q_a$ and thus of the harmonic functions $H_a$ the four-torus completely factorizes from the rest and has fixed volume/radii, while the $x^5$ circle with constant radius is fibered over the five-dimensional TS base space due the KK monopole.

\section{Rotating Topological Stars}
The goal of this section is to study solutions of minimal supergravity in five dimensions describing rotating topological solitons, to which we refer as \emph{Rotating Topological Stars} (or RTS for brevity). Our conventions are such that the bosonic action of minimal supergravity \cite{Gunaydin:1983bi} is given by
\begin{equation}
\label{EM-CS_action}
S\,=\,\frac{1}{2\kappa_5^2}\int \diff^5 x \sqrt{-g}\left(R-\frac{1}{4}F_{\mu\nu}F^{\mu\nu}-\frac{1}{12\sqrt{3}}\epsilon^{\mu\nu\rho\sigma\lambda}F_{\mu\nu}F_{\rho\sigma}A_{\lambda}\right)\ , 
\end{equation}
where $\epsilon^{01234}=-\frac{1}{\sqrt{-g}}$ and $F_{\mu\nu}=\partial_{\mu}A_{\nu}- \partial_{\nu}A_{\mu}$.

\subsection{The solution}

The starting point is a three-parameter family of solutions of \eqref{EM-CS_action} constructed by Compere, de Buyl, Jamsin and Virmani in \cite{Compere:2009zh} by using a solution-generating technique exploiting G$_2$ dualities arising from reductions from five to three dimensions. A solution in this family is obtained from a Kerr black string of mass $\mathtt{m}$ and rotation parameter $a$, by acting with a G$_2$ transformation parametrized by a duality phase $\d$. 
A more symmetric parametrization of the general solution can be obtained by introducing the following quantities,
\begin{equation}
r_b\,=\,2 \mathtt{m}\sinh^2\delta\,, \hspace{1cm} r_s\,=\, 2 \mathtt{m} \cosh^2\delta, \, \hspace{1cm}  a^2=a_s^2-a_b^2\, ,
\end{equation}
which are subject to the following constraint
\be
\label{constraint}
r_s\,a_b^2 \,=\, r_b\,a_s^2\, .
\ee
In terms of these parameters and the variables $\chi=\cos \theta$ and $r=r_{_\text{CdBJV}}+r_b$ the solution of \cite{Compere:2009zh} can be written as:
\begin{equation}
\label{RTS_solution}
\hspace{-7mm}
\begin{aligned}
\diff s^2\,=\,&-\left(1-\frac{r_s r}{\Sigma}+r_sr_b\,a_s^2\frac{\c^2}{\Sigma^2}\right) \diff t^2+\left(1-\frac{r_b r}{\Sigma}-r_sr_b\,a_b^2\frac{\c^2}{\Sigma^2}\right) \diff y^2 + \Sigma \left(\frac{\diff r^2}{\Delta}+\frac{\diff \c^2}{1-\c^2}\right) \\[1mm]
+&\frac{1-\c^2}{\S}\left[\left(r^2+a^2\right)^2-a^2\left(1-\c^2\right)\Delta+r_s r_b\, a^2\frac{ r^2 \left(1-\c^2\right)}{\Sigma}\right]\diff\phi^2 +2 \,r_sr_b\,a_sa_b\frac{\c^2}{\Sigma^2}\, \diff t \, \diff y\\[1mm]
-&2\,\frac{1-\c^2}{\S} \left[ r_s\,a_s \left(r-r_b\, a^2\frac{\c^2}{\Sigma}\right)\diff t \,\diff \phi-  r_b\,a_b \left(r-r_s \,a^2\frac{\c^2}{\Sigma}\right)\diff y \,\diff \phi\right]\,, \\[1mm]
A\,=\,& \sqrt{3 r_s r_b}\, \frac{\c}{\Sigma}\left(a_s \,\diff t-\left(r^2+a^2\right)\,\diff \phi-a_b\,\diff y\right)\, ,
\end{aligned}
\end{equation}
where
\begin{equation}
\Sigma\,=\,r^2+a^2\,\c^2\,, \hspace{1cm} \Delta=\left(r-r_s\right)(r-r_b)+a^2  \,.
\end{equation}

 \begin{figure}[t]
  \begin{minipage}{\textwidth}
    \centering
\includegraphics[width=0.8\textwidth]{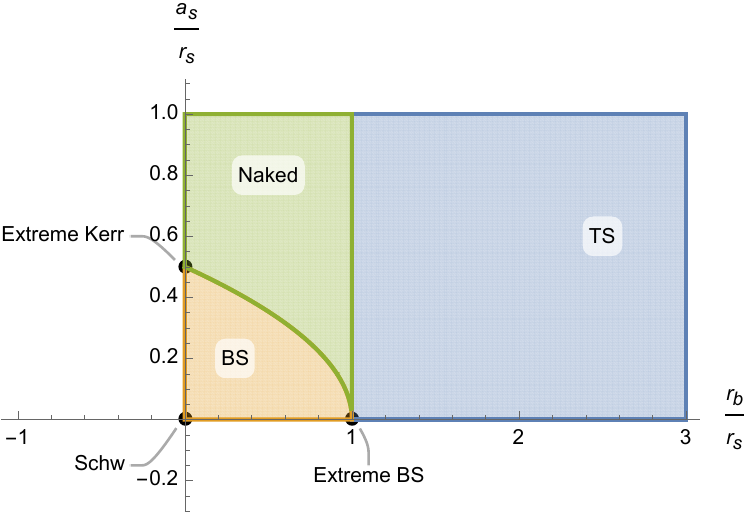}
    \caption{\it  The phase diagram for RTS's in terms of the dimensionless ratios $r_b/r_s$ \& $a_s/r_s$. The static case is recovered by projecting onto $t_2=0$, while the usual Kerr string case correponds to the $t_1=0$ line. The black string and naked singularity phases are separated by an arc of parabola setting the bound on the rotation parameter $a_s$ as to avoid over-rotation. The extreme black string plays the role of a triple point in the phase diagram. }
    \label{fig:phasediag}
  \end{minipage}
\end{figure}

Note that the $a\to 0$ limit of \eqref{RTS_solution} boils down to the static Horowitz-Strominger black string \cite{Horowitz:1991cd} in \eqref{TS_5d} when $r_s>r_b$ or to the Bah-Heidmann Topological Star \cite{Bah:2020ogh} when $r_b>r_s$. A first advantage of the parametrization we are using is that now the solution can be analytically continued for all values of $r_s$ and $r_b$, which is precisely what we need to do in order to describe a rotating topological soliton. As a bonus, it makes explicit the symmetry of the metric under the following transformations,
\be
\begin{array}{lll}
\left(t,y\right) \ \leftrightarrow \ \left(iy,it\right) \ , & \left(r_s,r_b\right) \ \leftrightarrow \ \left(r_b,r_s\right) \ , & \left(a_s,a_b\right) \ \leftrightarrow \ \left(ia_b,ia_s\right) \ ,
\end{array}
\label{s/b_flips}
\ee
which keep the metric and gauge field real.  As in the static case, the solution \eqref{RTS_solution} can be viewed as a superposition between a Lorentzian and a Euclidean Kerr solution, supported by a magnetic flux. Indeed, Lorentzian, Euclidean Kerr and the static black string/Topological Star solutions all appear as different limits of \eqref{RTS_solution}:
\begin{itemize}
\item $r_s \,=\, r_b \,=\, 0:~ \mathbb{R}^{1,4}$ written in oblate coordinates.
\item $r_b \,=\, a_b \,=\, 0$:  Kerr$_4\times S^1_y$ (Schw$_4\times S^1_y$ when $a_s=0$).
\item $r_s \,=\, a_s \,=\, 0$: EKerr$_4\times \mathbb{R}_t$  (ESchw$_4\times \mathbb{R}_t$ when $a_b=0$).
\item $a_s \,=\, a_b \,=\, 0, r_s\ge r_b$: Horowitz-Strominger black string \cite{Horowitz:1991cd}.
\item $a_s \,=\, a_b \,=\, 0, r_b> r_s$: static Bah-Heidmann Topological Star \cite{Bah:2020ogh}.
\end{itemize}
In what follows we consider the case in which none of the parameters $r_{s, b}$ and $a_{s, b}$ vanishes, and investigate the different phases of the solution. The latter fall into three classes, depending on whether space terminates on a curvature singularity (naked or surrounded by a horizon) or a smooth cap. Denoting by $r_\pm$ the roots of $\Delta(r)$, we find
\be
 \label{rpm}
r_\pm\,=\, \frac{r_b+r_s}{2} \,\pm\, \sqrt{ \left(\frac{r_b-r_s}{2}\right)^2 -a^2 }\,, 
\ee
recalling that 
\begin{equation}
a^2\,=\, a_s^2 - a_b^2\,=\, a_s^2\left(1-\frac{r_b}{r_s}\right)\,=\,-a_b^2\left(1-\frac{r_s}{r_b}\right)\quad {\rm or} \quad  a_s^2 \,=\, - { a^2 r_s \over r_b-r_s} \: , \: a_b^2 \,=\, - { a^2  r_b \over r_b-r_s} \quad .
\end{equation}
Then, we have that $a^2$ is positive for $r_s>r_b$. This corresponds to the case studied in \cite{Compere:2009zh}, in which space ends on a curvature singularity, either naked or surrounded by two horizons at $r_\pm$, according to the sign of the quantity under the square root. For $r_b> r_s$, instead, we show below that the geometry smoothly ends at $r_+$, which now represents a cap instead of an event horizon. Thus, this phase describes a horizonless topological soliton. A concise summary of the different phases is the following (see
figure \ref{fig:phasediag} and  section~\ref{subsec:phase_diagram} for a detailed discussion of the phase diagram): 
\begin{itemize}
\item  Naked singularity (NS): $r_s \,>\, r_b$\,, $a^2>(\ft{r_s-r_b}{2} )^2$.
\item Rotating black string (RBS): $r_s \,>\, r_b\,$, $0< a^2   \leq  (\ft{r_s-r_b}{2} )^2$. 
\item Rotating Topological Star (RTS): $r_b \,>\, r_s\,$, $a^2 <0$.
\item Extreme rotating black string (ERBS): $r_s \,=\, r_b$\,, $a^2=0$.
\end{itemize}
Because of the constraint \eqref{constraint}, the general solution has only three independent parameters, which can be taken to be $r_s, r_b \in [ 0,\infty)$, and one between $a^2,a_s^2 ,a_b^2  \in (-\infty,\infty)$. Moreover, in the BS and RTS phases, where $r_\pm$ are real, the general solution can be alternatively described in terms of $r_s, r_b$ and $r_+$. The explicit expression of the rotation parameters in terms of these variables is
 \be
a_{s, b}\,=\,\sqrt{ {\left(r_+ - r_b\right)\left(r_+ - r_s\right) r_{s, b}\over r_b-r_s}} \, , \hspace{1cm} a^2\,=\, -\left(r_+ - r_b\right)\left(r_+ - r_s\right) \, . 
\ee

Let us study in more detail the RBS and RTS phases and comment on ERBS at the very end. 
 
\subsection{Rotating black strings}  

This case has been already analyzed in \cite{Compere:2009zh}, so we keep the discussion brief. Our goal here is to emphasize those aspects which will be useful when studying the solution in the topological soliton regime. As already stated, the rotating black string regime corresponds to
\begin{equation}
r_s \,>\, r_b\,, \hspace{1cm}   0< a^2   \leq  (\ft{r_s-r_b}{2} )^2\, ,
\end{equation}
which together with \eqref{rpm} implies
\begin{equation}
r_s>r_+ >r_b\, .
\end{equation}
The hypersurface $r=r_+$ is a Killing horizon for the following Killing vector,
\begin{equation}
\label{eq:gen_hor}
\zeta\,=\, \partial_t  + \zeta^{\phi} \,\partial_{\phi}+\zeta^y \,\partial_y  ,
\end{equation}
where
\be 
\label{ephi0}
\zeta^{\phi} \,=\, \ \frac{r_s{-}r_+}{a_s \,r_s}\,=\,  \sqrt{\frac{\left(r_s{-}r_b\right)\left(r_s{-}r_+\right)}{r_s^3\left(r_+{-}r_b\right)}}\,, \hspace{2.5mm}\quad  \zeta^y\,=\,-  \frac{a_b\,r_b}{a_s\,r_s}\frac{r_s{-}r_+}{r_+{-}r_b} \,=\,{-} \left(\frac{r_b}{r_s}\right)^{3/2} \frac{r_s{-}r_+}{r_+{-}r_b} \, .
\ee 
To study the geometry near the horizon, it is convenient to introduce the coordinates  
\begin{equation}
\tilde t\,=\, t\,, \hspace{1cm} {\tilde \phi}\,=\, {\tilde \phi}- \zeta^{\phi}\, t\,, \hspace{1cm} \tilde y\,=\, y-\zeta^{y}  \, t\, ,   \hspace{1cm} \varrho^2\equiv 4\left(r-r_+\right) \,,
\end{equation}
in terms of which the above Killing vector becomes simply $\zeta\,=\, \partial_{\tilde t}$. The metric at leading order in $\varrho$ is given by\footnote{We ignore the mixed components $g_{\tilde t \tilde \phi}$ and $g_{\tilde t \tilde y}$, which vanish in the $\varrho\to0$ limit.}
\begin{equation}
\label{eq:near_horizon_metric}
\diff s^2\,=\, \frac{\Sigma_+}{2r_+-r_b-r_s}\left(-\kappa^2\varrho^2 \diff {\tilde t}^2+\diff \varrho^2\right)+ \diff s^2_{\cal H}\,, 
\end{equation}
where $\Sigma_+ \equiv \Sigma \left(r_+\right)$,
\begin{equation}
\kappa^2\,=\, \frac{(r_s-r_b)\left(2r_+-r_b-r_s\right)^2}{4\left(r_+-r_b\right)^2 r_s^3}\,
\end{equation}
is the square of the surface gravity and 
\begin{equation}
\begin{aligned}\label{horiz}
\diff s^2_{\cal H}\,=\,&\frac{\Sigma_+\,\diff \chi^2}{1-\chi^2}+ \frac{\left(1-\chi^2\right)\, F_{\cal H}\left(\chi\right)}{\Sigma_+}\,\diff {\tilde\phi}^2- \left(1-\frac{G_{\cal H}\left(\chi\right)}{\left(r_s-r_b\right)\Sigma_+^2}\right)\diff {\tilde y}^2\\[1mm]
&+\frac{2 r_b \left(1-\chi^2\right)}{\Sigma_+^2}\sqrt{\frac{r_b \left(r_+-r_b\right)\left(r_+-r_s\right)}{r_b-r_s}}\left[r_+^3-\left(r_+ - r_b\right)\left(r_+ - r_s\right) \chi^2\right]\diff {\tilde y} \,\diff {\tilde \phi}\, 
\end{aligned}
\end{equation}
is the induced metric at the horizon. The functions ${F}_{\cal H}$ and $G_{\cal H}$ are explicitly given by
\begin{equation}
\begin{aligned}
F_{\cal H}(\chi)\,=\,&\left(r_+ \, r_b + r_+ \, r_s-r_s \,r_b\right)^2-\frac{r_s \,r_b \,r_+^2 \, \left(r_+ - r_b\right)\left(r_+ - r_s\right)  \left(1-\chi^2\right)}{\Sigma_+} ,\\[1mm]
G_{\cal H}(\chi)\,=\, &r_+^3r_b\left(r_s-r_b\right)- r_b \left(r_+ - r_b\right)\left(r_+ - r_s\right) \left(r_+ \, r_b - r_+ \, r_s-r_s \,r_b\right)\chi^2\,.
\end{aligned}
\end{equation}
The near-horizon metric \eqref{eq:near_horizon_metric} is Rindler${}_{2}\times S^2\times S^1$. The Rindler factor (which characterizes the near-horizon geometry of non-extremal black holes) is parametrized by the coordinates $t$ and $\varrho$; slices of the horizon at constant $t$ have $S^2\times S^1$ topology, with the $S^1$ being parametrized by the coordinate $y$, which is assumed to be compact. The $S^1$ never shrinks outside the horizon, and has finite size at infinity, where the solution tends to ${\mathbb R}^{1, 3}\times S^1$. We conclude by observing the following property, 
 \begin{equation}
F_{\cal H}\left(\pm 1\right)\,=\, \Sigma^2_+\left(\pm 1\right)\, ,
\end{equation}
which ensures regularity of the metric \eqref{horiz} at the poles of the $S^2$. 

\subsection{Rotating Topological Stars}
\label{sec:RTS}

Next, we consider the topological soliton regime,
\be
r_b \,>\, r_s   \hspace{1cm}\Rightarrow \hspace{1cm}  a^2 <0\, .
\ee
Taking into account \eqref{rpm}, the above implies
\begin{equation}
\label{eq:regime_TS}
r_+>r_b >r_s\, .
\end{equation}
Contrarily to the RBS case, the hypersurface $r=r_+$ no longer corresponds to an event horizon, but rather to the locus where an $S^1$ shrinks to zero size, as we now show.

\paragraph{Regularity at the cap.} Consider the Killing vector
\begin{equation}
\label{eq:Killing_cap}
\xi\,= \partial_y  + \xi^{\phi} \,\partial_{\phi}+\xi^{t}\, \partial_t\,  ,
\end{equation}
where
\be
 \label{ephi}
\xi^{\phi} \,=\, - \frac{r_+-r_b}{a_b \,r_b}\,=\,- \sqrt{ (r_+-r_b) (r_b-r_s)\over r_b^3 (r_+-r_s)} \,, \hspace{2.5mm}  \xi^t \,=\, \ \frac{a_s\,r_s}{a_b\,r_b} \,\frac{r_+-r_b}{r_+-r_s} \,=\,\frac{ r_s^{3/2} (r_+-r_b)}{ r_b^{3/2} (r_+-r_s)}\,.
\ee 
This is proportional to the generator of the horizon in \eqref{eq:gen_hor} and \eqref{ephi0},\footnote{We note, however, that the proportionality constant vanishes in the static limit, where we already know that they are not proportional to each other.} which makes it evident that its norm also vanishes at $r=r_+$,
\begin{equation}
g_{\mu\nu}\xi^{\mu}\xi^{\nu}|_{r_+}\,=\, 0\, .
\end{equation}
Contrarily to the black string case, $\xi$ is spacelike rather than timelike near $r\,=\,r_{+}$. This fact, together with the symmetry \eqref{s/b_flips}, strongly motivates us to introduce the following coordinates,
 \begin{equation}
 \label{tpiden}
t'\,=\, t-\xi^t \,y\,, \hspace{1cm}  \phi' \,=\, \phi -\xi^{\phi}\, y\,,  \hspace{1cm} y'\,=\,  \, y \, , \hspace{1cm} \varrho^2\,=\,4(r-r_+) \, ,
\end{equation}
which are adapted to the isometry generated by $\xi$. In terms of them, this Killing vector simply becomes $\xi\,=\,\partial_{y'}$ and the metric near the cap, at leading order in $\varrho$, becomes
  \begin{equation}
  \label{cap_metric}
\diff s^2\,=\, \frac{\Sigma_+}{2r_+-r_b-r_s}\left(\diff \varrho^2+ \ft{\varrho^2 }{\gamma^2}\, \, \diff  y'^2\right)+\diff s^2_{\rm cap} \, ,
\end{equation}
with
\be
\gamma =\frac{2\,r_b^{3/2} \left(r_+-r_s\right)} {\sqrt{r_b-r_s}\left(2r_+-r_b-r_s\right)} \,,
\label{gammary0}
\ee
and 
 \begin{equation}
\label{eq:induced_metric_cap}
\begin{aligned}
\diff s^2_{\rm{cap}}\,=\,&-\left(1-\frac{G_{\rm cap}(\chi)}{\left(r_b-r_s\right)\Sigma_+^2}\right){\diff t'}^2+\frac{\Sigma_+\, \diff \chi^2}{1-\chi^2} +\frac{\left(1-\chi^2\right){F}_{\rm cap}(\chi)}{\Sigma_+} {\diff \phi'}^2\\[1mm]
&-\frac{2r_s \left(1-\chi^2\right)}{\Sigma_+^2} \sqrt{\frac{r_s \left(r_+ - r_b\right)\left(r_+ - r_s\right) }{r_b-r_s}}\left[r_+^3-\left(r_+-r_b\right)^2\left(r_+-r_s\right)\chi^2\right]\, \diff t'\, \diff \phi'\, ,
\end{aligned}
\end{equation}
is the induced metric at the cap, which has ${\mathbb R}\times S^2$ topology. The functions $F_{\rm cap}(\chi)$ and $G_{\rm cap}(\chi)$ are given by 
\begin{equation}
\begin{aligned}
{F}_{\rm cap}\left(\chi\right)\,=\,& F_{\cal H}(\chi)\,=\,\left(r_+ \, r_s+r_+ \, r_b-r_s \,r_b\right)^2-\frac{r_s \,r_b \,r_+^2 \,\left(r_+-r_b\right)\left(r_+-r_s\right) \left(1-\chi^2\right)}{\Sigma_+}\,, \\[1mm]
G_{\rm {cap}}(\chi)\,=\, &r_+^3 \, r_s\left(r_b-r_s\right)+r_s \left(r_b-r_+\right)\left(r_+-r_s\right) \left(r_+ \, r_b- r_+\, r_s+r_b \,r_s\right)\chi^2\,.
\end{aligned}
\end{equation}
and coordinates are subjected to the global identifications
 \be
 \label{eq:periodic_identifications}
 (\phi',y')\sim  (\phi' +2\pi ,y') \sim (\phi', y'+2\pi R_y) \, ,
 \ee
 The two dimensional  metric  along the $(\rho,y')$ plane is flat near the cap, and it exhibits  in general a conical singularity at the origin with excess/defect angle $2\pi(\ft{\gamma}{R_y}-1)$. 
 Although a general choice may lead to interesting scenarios as well, see for example  \cite{Balasubramanian:2000rt} for applications of excess/defect conical singularities to black hole formation, here we restrict ourselves to the regular case,  so we take
   \be
   R_y=\gamma=\frac{2\,r_b^{3/2} \left(r_+-r_s\right)} {\sqrt{r_b-r_s}\left(2r_+-r_b-r_s\right)}\label{gammary}
   \ee

 \paragraph{The asymptotic geometry.} Having analyzed the geometry near $r=r_+$, let us consider now the geometry at infinity. The metric \eqref{RTS_solution} always asymptotes to flat spacetime,
\begin{equation}
\diff s^2\, =\,-\diff t^2 + \diff r^2 +r^2 \left(\diff \theta^2+\sin^2\theta \,\diff \phi^2\right)+\diff y^2\,+ \dots\,,  \label{flat}
\end{equation}
where the dots denote subleading terms in the $r\to \infty$ limit. Contrarily to what happens for the RBS, the coordinates $t, \phi, y$ now satisfy twisted periodic identifications, inherited from \eqref{eq:periodic_identifications},
\begin{equation}
\label{ident3}
\left(t, \phi, y \right)\sim \left(t, \phi+2\pi, y\right)\, \sim \left(t+ 2\pi R_y\,\xi^t ,\phi+2\pi  R_y\,\xi^{\phi} , y+2\pi R_y \right) \, ,
\end{equation}
involving also the time coordinate. This can be remedied by boosting the solution along the $y$ direction with velocity $\xi^t$\footnote{We notice that the same happens for the solitons studied in \cite{Giusto:2004ip}, where the authors first introduce an arbitrary boost parameter and then fix it by imposing the time coordinate does not shift. We thank Enrico Turetta for pointing to us the similarity between our analysis and that of \cite{Giusto:2004ip}, and to Stefano Giusto for correspondence on this point.} ,
\begin{equation}
\hat t\,=\, \frac{t-\xi^t \,y}{\sqrt{1-\xi^t{}^2}}\,, \hspace{1cm }\hat y\,=\, \frac{y-\xi^t\, t}{\sqrt{1-\xi^t{}^2}} \, 
\end{equation}
that leaves invariant the asymptotic metric \eqref{flat}. Notice that $\xi^t$, given in \eqref{ephi}, satisfies $0<\xi^t<1$ in the soliton regime \eqref{eq:regime_TS}.
 The identifications \eqref{ident3} now become\begin{equation}
\label{eq:ident_hatted_coord}
\left(\phi, \hat y\right) \sim \left(\phi+2\pi, \hat y\right)\, \sim \left( \phi + 2\pi R_y \,\xi^{\phi}, \hat y + 2\pi R_y \sqrt{1-\xi^t{}^2}\right) \, .
\end{equation}
Moreover,   in the new coordinates the Killing vector \eqref{eq:Killing_cap}, whose orbits generate the $S^1$ shrinking at $r=r_+$, can be written as a linear combination of the two  compact isometries
\begin{equation}
\label{eq:xi}
\xi\,=\, \partial_{y'}=\sqrt{1-\xi^t{}^2}\, \partial_{\hat y}+ \xi^{\phi}\, \partial_{\phi}\, .
\end{equation}

\begin{figure}[h]
\centering
\includegraphics[scale=1.2]{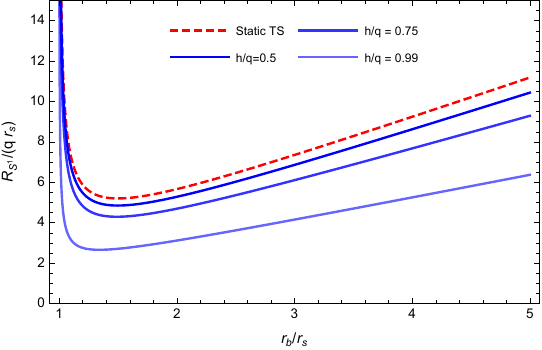}
\caption{\emph{$\frac{R_{S^1}}{q \,r_s}$ as a function of $\frac{r_b}{r_s}$ for different values of $\frac{h}{q}$.}}
\label{fig:RS1}
\end{figure}

The identifications \eqref{eq:ident_hatted_coord} become untwisted if the product $R_{y}\,\xi^{\phi}$ is an integer number.  Unfortunately, this condition does not admit real solutions in the rotating case. The best can be done, is to assume that the product is a fractional number, so we require that
\be
\label{ryn}
R_y \,  \xi^{\phi}\,=\,- {h\over q } \,,
\ee
where $h$ and $q$ are co-prime integers such that $q>h\geq 0$. This condition ensures that we come back to the same point after going $q$ times along the $\hat y$-circle. It is also equivalent to demanding that the orbits of $\xi$ are not dense in the torus generated by $\partial_{\hat y}$ and $\partial_{\phi}$ \cite{Gibbons:1979xm}. 
As a result, the asymptotic geometry is given by the orbifold 
 $(\mathbb{R}^{1,3}\times S^1)/ \mathbb{Z}_q$, where  $\mathbb{Z}_q$ acts freely as a rotation on $\mathbb{R}^{1,3}$ accompanied with a shift or order $q$ along $S^1$. 
 More precisely, the action of ${\mathbb Z}_{q}$ is such that
 \be
\left( \phi, \hat y\right) \to \left( \phi + \frac{2\pi h}{q},  \hat y + { 2\pi  R_{S^1}  \over q} \right) \,, 
 \ee
 where 
\be
\label{rs1} 
R_{S^1}\,=\, q R_y \sqrt{1-\xi^t{}^2} \,.
 \ee
Alternatively, one can relax the condition $R_y=\gamma$ and allows for a conical singularity at the cap. For this choice, using (\ref{gammary0}) and (\ref{ryn}) one finds in general an excess/defect angle ratio
      \be
  {R_y\over \gamma}={h\over 2q} \left[ 1+{r_+-r_s\over r_+-r_b} \right] \label{rexcess}
  \ee 
  In particular for $q=1$, where the orbifold acts trivially and  five dimensional geometry factorizes as $\mathbb{R}^{3,1}\times S^1$, the right hand side of (\ref{rexcess}) is always bigger than one in the RTS regime $r_+>r_b>r_s$ leading to an excess conical singularity at the cap. In the remaining of this work, we will always assume that the cap is smooth, so the 
  ratio in (\ref{rexcess}) is simply one.   We notice that this requirement \eqref{ryn} boils down to a  quantization condition on the angular momentum,
 \be
 \label{eq:quantization}
 a^2\,=\, - {h^2(r_b-r_s)^2\over 4\left(q^2- h^2\right)}\, , \hspace{1cm} h<q\, .
  \ee
Finally, plugging this into \eqref{rs1}, one finds the following constraint relating $R_{S^1}$, $r_s$ and $r_b$:
\be
\label{eq:RS1}
R_{S^1}^2 \,=\, {c_+\, r_b^3 -c_- r_s^3\over r_b-r_s}\,,
\ee
where
\be
c_\pm \,=\,2q^2-h^2\pm 2\sqrt{q^2-h^2}\, .
\ee
In the static case $h=0$, \eqref{eq:RS1} correctly reduces to the result in \cite{Bah:2020ogh}, namely
\begin{equation}
R^2_{S^1} \big|_{h= 0}= \, \frac{4 \, r_b^3}{r_b-r_s}\, .
\end{equation}
We notice that  $R_{S^1}|_{h=0}$  has an absolute minimum at $r_b\,=\, 3 r_s/2$, where it takes the value $R_{S^1}/r_s \,=\, 3\sqrt{3}$. The existence of a lower bound for $R_{S^1}$ persists in the rotating case, as we show in Fig.~\ref{fig:RS1}. This tells us that there is no net separation of scales between the size of the $S^1$ and the would-be horizon scale, $r_+$.
As in the static case, the RTS solution is genuinely five-dimensional.

 Finally, we notice that, as in the static case,  once the asymptotic geometry is fixed, i.e. given $R_{S^1}, h, q$; the solution is given in terms of a  single continuous parameter, let us say  $r_s$ or the mass.  All other quantities $a$, $r_b$, $r_c$   can be determined in terms of $\{ R_{S^1}, h/q, r_s\}$  by \eqref{eq:quantization} and \eqref{eq:RS1}. This is equivalent to say that the magnetic charge and angular momentum are fixed in terms of the asymptotic boundary data and the mass.

\paragraph{Absence of an ergoregion.} In pure General Relativity, spinning compact objects featuring an ergoregion without an event horizon are known to be unstable \cite{Friedman:1978ygc}, since physical negative-energy states existing inside the ergoregion are energetically favorable to cascade toward even more negative states.  As shown in \cite{Cardoso:2005gj}, these `ergoregion instabilities' affect other non-supersymmetric smooth horizonless geometries,\footnote{Even the supersymmetric ones, which have an `evanescent ergosurface' \cite{Gibbons:2013tqa}, have been argued to be non-linearly unstable \cite{Eperon:2016cdd}.} such as JMaRT \cite{Jejjala:2005yu}, that also suffer from charge instabilities \cite{Bianchi:2023rlt}. 
  
Here we show that no potential issue of this kind may take place in our solutions since the Killing vector $\partial_{\hat t}$ is everywhere timelike. Its norm, given by the $\hat t \hat t$ component of the metric, reads
\begin{equation}
\label{eq:no_ergoregion}
-g_{\hat t \hat t}\, =\,  1+ \frac{\xi^t{}^2 r_b-r_s}{1-\xi^t{}^2 } \, \frac{r}{\Sigma}+ \frac{\left(a_s-\xi^t a_b\right)^2}{1-\xi^t{}^2}  \, \frac{r_s r_b  \chi^2}{\Sigma^2}\, .
\end{equation}
The last term is manifestly positive, so we just have to show that the remaining ones be positive as well. To this aim, we recall that in the soliton regime $r\ge r_+$ and 
$a^2<0$,  which implies that $0<\Sigma<r^2$, and $\xi^t<1$. Then,
\begin{equation}
1+ \frac{\xi^t{}^2 r_b-r_s}{1-\xi^t{}^2 } \, \frac{r}{\Sigma} > 1+ \frac{\xi^t{}^2 r_b-r_s}{1-\xi^t{}^2 } \, \frac{1}{r} >\left(1-\xi^t{}^2\right)\frac{r-r_b}{r}>0\,,
\end{equation}
where we used the fact that space ends at $r=r_+>r_b>r_s$. Thus, we conclude that the would-be ergoregion is not part of the RTS spacetime.

\paragraph{Absence of closed timelike curves (CTC).} In order to check the absence of CTC in our geometry, we follow \cite{Elvang:2004ds}, where it is argued that the positive-definiteness of the induced metric on a Cauchy slice at constant $\hat t$ ensures the absence of CTC in a region where $\partial_{\hat t}$ is timelike.\footnote{As a matter of fact, this condition guarantees the absence of closed causal curves, \cite{Elvang:2004ds}.} We emphasize that although the authors apply this in a supersymmetric context, the argument holds whenever the latter condition is satisfied. As we showed in \eqref{eq:no_ergoregion}, in our geometry $\partial_{\hat t}$ is always timelike. Hence, we just have to check that the induced metric on $\hat t \,=\, \rm{constant}$ hypersurface is positive definite. 

Given the form of the metric, this reduces to a two-dimensional problem, since the $r-\chi$ part is automatically positive-definite. The relevant two-dimensional metric is then the following,
\begin{equation}
\diff s^2_{\phi,\hat y}\,=\, g_{\hat y \hat y}\left(\diff \hat y + \frac{g_{\hat y\phi}}{g_{\hat y \hat y}}\diff \phi\right)^2 + \frac{\cal D}{g_{\hat y \hat y}}\, {\diff \phi}^2\, ,
\end{equation}
where $\cal D$ is the determinant. Thus, the metric is positive-definite if 
\begin{equation}
g_{\hat y \hat y}>0\,, \hspace{1cm} {\cal D}>0\, .
\end{equation}
In order to demonstrate the strict inequality one has to stay away from the poles $\chi\,=\,\pm 1$, where we already know that regularity imposes that both $\cal D$ and $g_{\hat y \phi}$ vanish. Similarly, we know that $g_{\hat y \hat y}$ must vanish at the poles of the $S^2$ at $r=r_+$, since these correspond to the fixed points of the isometry $\partial_{\hat y}$. This can be deduced from \eqref{eq:xi}, where we showed that $\partial_{\hat y}$ is a linear combination of $\partial_{y'}$ and $\partial_{\phi}$ which vanish, respectively, at $r=r_+$ and $\chi=\pm 1$.

A sketch of the proof is the following. The expressions for $\cal D$ and $g_{\hat y \hat y}$ are of the form,
\begin{equation}
{\cal D}\,=\, \frac{1-\chi^2}{r_s^2 \left(1-\xi^t{}^2\right)\Sigma} \left({\cal D}_{(0)}+\left(1-\chi^2\right) {\cal D}_{(1)}\right)\, ,
\end{equation}
and 
\begin{equation}
g_{\hat y \hat y}\,=\,  \frac{1}{r_s^2 \left(1-\left(\xi^t\right)^2\right)\Sigma^2}\left[{\cal G}_{(0)}+{\cal G}_{(1)}\left(1-\chi^2\right)+{\cal G}_{(2)}\left(1-\chi^2\right)^2 \right]\,, 
\end{equation}
where ${\cal D}_{(0)}$, ${\cal D}_{(1)}$, ${\cal G}_{(0)}$, ${\cal G}_{(1)}$ and ${\cal G}_{(2)}$ are functions of $r$, whose explicit expression is not displayed as it is somewhat lengthy and unilluminating. The important point is that one can check that all of them are positive (except at the fixed points of $\partial_{\hat y}$). This ensures there are no CTC in the RTS geometry.

\paragraph{GL instabilities.} Black string and extended solutions are known to suffer from Gregory-Laflamme (GL) instabilities \cite{Gregory:1993vy} if the Kaluza-Klein radius is not small enough. This is certainly the case of black string solutions along the $y$-axis ($r_b=0$)  of the phase diagram \ref{fig:phasediag}. On the other hand, static solutions \cite{Horowitz:1991cd, Bah:2020ogh}, although generically unstable, have been shown to be free of GL instabilities if falling in the segment  $\ft12 <\ft{r_b}{r_s}<2$ along the $x$-axis \cite{Bena:2024hoh, Miyamoto:2007mh}.  This suggests that rotating solutions with $\ft12 <\ft{r_b}{r_s}<2$ should be GL stable when the rotation parameter is small enough.  A  rigorous claim, however, would require a calculation of QNM frequencies of all linear perturbations of the rotating metric, a challenging task that can be addressed by with recently proposed methods based on the  duality with ${\cal N}=2$ supersymmetric gauge theories \cite{Aminov:2020yma,Bianchi:2021xpr,Bianchi:2021mft,Bonelli:2021uvf,Bonelli:2022ten,Bianchi:2022qph,Consoli:2022eey,Bautista:2023kns,Bianchi:2024zgn,DiRusso:2024rgi,Cipriani:2025ikx}, but is beyond the scope of our present work.

\subsection{Extreme rotating black strings}  
We notice that the NS and RTS  regions are divided by a continuous line of solutions (see figure~\ref{fig:phasediag}) corresponding to the choice
   \be
   r_+\,=\, r_b \,=\, r_s \,, \, \hspace{1cm} a_b\,=\,a_s  \,, \hspace{1cm}  a^2\,=\, 0 \, ,\nn\\
   \ee 
whereby the functions $\Delta$ and $\Sigma$ simply reduce to
   \be
   \Delta\,=\,(r-r_s)^2 \, , \hspace{1cm} \Sigma\,=\,r^2 \ .
   \ee
The corresponding form of the 5d metric becomes
\be
\diff s^2 \,=\, \diff s_{\mathrm{EBS}}^2 \,-\, a_s^2r_s^2 \frac{\c^2}{r^4}(\diff t-\diff y)^2 \,-\, 2a_s\,r_s\frac{1-\c^2}{r}(\diff t-\diff y)\diff \phi\ ,
\ee
which could be understood as a one-parameter deformation of the extreme static black string solution, whose line element we denote by $\diff s_{\mathrm{EBS}}^2$:
\be
\diff s_{\mathrm{EBS}}^2 \,=\, f_s \left(-\diff t^2+\diff y^2\right) \,+\, \frac{\diff r^2}{f_s^2}\,+\, r^2 \diff s_{S^2}^2 \ ,
\ee
with $f_s \equiv \left(1-\dfrac{r_s}{r}\right)$. Introducing the coordinates
\begin{equation}
u\,=\, t-y\,, \hspace{1cm} v\,=\, t+y\,, 
\end{equation}
the metric and gauge field can be brought into the form
\begin{equation}
\begin{aligned}
\diff s^2\,=\,& -f_s \,\diff u \left(\diff v + \frac{a_s^2 r_s^2 \chi^2}{r^4f_s}\diff u + \frac{2 a_s r_s\left(1-\chi^2\right)}{r f_s} \, \diff \phi \right) \,+\, \frac{\diff r^2}{f_s^2}\,+\, r^2 \diff s_{S^2}^2 \, ,\\[1mm]
A\,=\,& -\sqrt{3 r_s r_b}\, \chi\left(\diff\phi - \frac{a_s}{r^2}\, \diff u\right)\, .
\end{aligned}
\end{equation}
This solution possesses mass and angular momentum and exhibits a curvature singularity at $r=0$.

\subsection{Five-dimensional charges}
\label{sec:5dcharges}

We are now ready to compute the five-dimensional charges  of the BS and RTS  solutions. These are given by the general formulae:
\be
{\cal M} \, = \, {3\over  4\kappa_5^2} \int\limits_{\Sigma_\infty} \star \diff K^{\sharp}_{(t)}\, ,\hspace{1cm}
{\cal J} \, = \, {1\over 2 \kappa_5^2} \int\limits_{\Sigma_\infty} \star \diff K^{\sharp}_{(\phi)}   \, , \hspace{1cm} 
{\cal P} \, = \, {1\over 2 \kappa_5^2} \int\limits_{\Sigma_\infty} \star \diff K^{\sharp}_{(y)}  \, , 
\label{charges5D}
\ee
where $K^{\sharp}_{(i)}=g_{i \mu} \, \diff x^\mu$ are the one-forms associated to the three Killing vectors $K_{(i)}$ generating the isometries of the solutions, and $\Sigma_\infty$ a spatial cross-section of the asymptotic boundary.  This is summarized in the table below.
   \[
\begin{array}{c|c|c|c|c}
 & ~~K_{(t)}~~ &  ~~K_{(\phi)} ~~& ~~K_{(y)}  ~~& \Sigma_{\infty} \\
 \hline
 {\rm RBS} & \partial_{t} & \partial_{\phi} &  \partial_{y} &  S^2 \times S^1  \\ 
{\rm RTS} & \partial_{\hat{t}} & \partial_{\phi} &  \partial_{\hat{y}} &  \left(S^2 \times S^1\right)/\mathbb{Z}_q   \\  
\end{array}
\]

\paragraph{Charges for the rotating black string.} The mass and angular momentum of the RBS can be extracted from the asymptotics of the $g_{tt}$ and $g_{t\phi}$ components of the metric. One finds
\be
\begin{aligned}
{\cal M}\,=\, & \frac{3V_{S^2\times S^1}}{4\kappa^2_5}\lim_{r\to \infty} \left(-r^2\, \partial_rg_{tt}\right)\,=\,\frac{3V_{S^2\times S^1}}{4\kappa^2_5} \, r_s\,,\\[1mm]
{\cal J}\,= \,& \frac{V_{S^2\times S^1}}{2\kappa^2_5}\lim_{r\to \infty} \left( r^2\, {\partial_rg_{t\phi} \over 1-\chi^2} \right)\,=\,  \frac{V_{S^2\times S^1}}{2\kappa_5^2} 2 a_s r_s\, ,
\end{aligned}
\ee
where 
\begin{equation}
V_{S^2\times S^1}\,=\, 8\pi^2 R_{S^1}\, .
\end{equation}
The momentum charge $ {\cal P}$ vanishes since $g_{ty}\,=\,{\cal O}\left(1/r^4\right)$.

\paragraph{Charges for the Rotating Topological Star.} Although the form of the solution is the same, the expressions for the charges of the topological soliton differ from the black string for two reasons. The first is the fact that the boundary is different, as we have explained in section~\ref{sec:RTS}. The second is the boost we have performed along the $y$ direction. This will give rise to a non-vanishing momentum charge ${\cal P}$, as can be seen from the asymptotic expansion of the metric:
\bea
\label{ds25inf}
\diff s^2  &\,=\,& - \left(1- \frac{r_s-\xi^t{}^2 r_b}{1-\xi^t{}^2} \,\frac{1}{r}  \right)\, \diff \hat{t}^2 + \left(1- \frac{r_b-\xi^t{}^2 r_s}{1-\xi^t{}^2} \,\frac{1}{r}  \right)\, \diff \hat{y}^2+ \diff r^2 + r^2 \diff s^2_{S^2}  \\[1mm]
&&- \frac{2\xi^t\left(r_b{-}r_s\right)}{1-\xi^t{}^2} \frac{1}{r}\, \diff\hat{t} \,\diff\hat{y}+2 \frac{1-\chi^2}{r}\left[\frac{a_b r_b \xi^t-a_s r_s}{\sqrt{1- \xi^t{}^2}}\, \diff \hat{t} \,\diff \phi {+} {a_b r_b{-}a_s r_s \xi^t\over  \sqrt{1{-}\xi^t{}^2} }\diff\hat{y} \,\diff\phi\right] +\ldots \,  \nn
\eea
where 
\begin{equation}
\diff s^2_{S^2}=\frac{\diff \chi^2}{1-\chi^2}+ \left(1-\chi^2\right) \,\diff \phi^2\,,
\end{equation}
 is the metric of the round $S^2$ and the dots denote $1/r^2$-corrections.  From \eqref{charges5D} one finds:
\be
\label{eq:5dchargesRTS}
\begin{aligned}
{\cal M}\, =\,& \frac{3V_{S^2\times S^1 }}{4 q \kappa^2_5}\lim_{r\to \infty} \left(-r^2\, \partial_rg_{\hat{t} \hat{t} }\right)\,=\frac{3V_{S^2\times S^1}}{4q\k_5^2}  \frac{r_s-\xi^t{}^2 r_b}{ 1-\xi^t{}^2}\,,  \\[1mm]
{\cal J}\, =  \,& \frac{V_{S^2\times S^1}}{2q \kappa^2_5}\lim_{r\to \infty} \left( r^2\, {\partial_rg_{ \hat{t}\phi} \over 1-\chi^2} \right)\,=\, \frac{ V_{S^2\times S^1} }{2q\k_5^2} \, {a_s \,r_s{-}a_b \,r_b \,\xi^t \over  \sqrt{1{-}\xi^t{}^2} }\,,\\[1mm]
{\cal P}\, =& \,\frac{V_{S^2\times S^1}}{2q \kappa^2_5}\lim_{r\to \infty} \left( r^2\, {\partial_rg_{ \hat{t}\hat{y} } \over 1-\chi^2} \right)\,=\,   
 \frac{V_{S^2\times S^1}}{q\k_5^2} \, {\left(r_b{-}r_s\right)\xi^t \over 1{-}\xi^t{}^2}\, .
 \end{aligned}
\ee
We conclude showing explicitly that the mass is positive, as it is not manifest from the above expression. However, using \eqref{ephi} we have that 
\begin{equation}
r_s-\xi^t{}^2 r_b\,=\,\frac{r_s}{\left(r_+-r_s\right)^2}\left[r_b^2 \left(r_+-r_s\right)^2 -r_s^2 \left(r-r_b\right)^2\right]>0\,, 
\end{equation}
since $r_+>r_b>r_s>0$.

\subsection{Four dimensional charges}

In this section we perform the dimensional reduction of the RTS solution down to four dimensions and compute its charges. There are two natural choices for the compactifications: reducing along $\partial_{y'}$ or along $\partial_{\hat{y}}$. The former choice leads to a magnetic flux tube background generalizing Melvin solution of Einstein-Maxwell theory \cite{Dowker:1995gb}\footnote{We thank Pierre Heidmann for the references and detailed discussions on this point.}. Here we focus on the reduction along $\partial_{\hat{y}}$ leading to an asymptotically flat four-dimensional space instead. We remind that the five-dimensional asymptotic  geometry has the topology $(\mathbb{R}^{1,3}\times S^1)/ \mathbb{Z}_q$, so non-spherical symmetric modes mix with KK momenta along the $S^1$. In addition, as in the static case, the radius of the circle varies along the spacetime and shrinks to zero size at $r=r_+$ where the entire four-dimensional picture beaks down. The RTS solution is genuinely five-dimensional. Bearing this in mind, we can still find an asymptotically flat  four-dimensional solution by reducing 
the RTS geometry along $\hat{y}$, and compute its charges.  More precisely, we start from the reduction ansatz:
\be
\begin{array}{lcl}
\diff s^2 &=& e^{\varphi} \,\diff s_{4}^2 \,+\, e^{-2\varphi}\,\left(\diff \hat{y}+\mathcal{B}\right)^2 \ , \\[2mm]
A &=& \mathcal{A}\,+\, \a \, \left(\diff \hat{y}+\mathcal{B}\right) \ ,
\end{array}
\label{KK_Ansatz}
\ee
with $\mathcal{A}$, $\mathcal{B}$ some one-forms in 4d. 
Assuming that all five-dimensional fields be $\hat{y}$-independent, the five-dimensional action \eqref{EM-CS_action} reduces to 
\begin{equation}
\begin{aligned}
S_4\,=\,\frac{1}{2\k_4^2}\int \diff^4 x \sqrt{-g}\,\bigg[& R_{4}-\frac{3}{2}\left(\partial\varphi\right)^2-\frac{e^{2\varphi}}{2}\left(\partial\a\right)^2-\frac{e^{-\varphi}}{4}|F_{(2)}|^2-\frac{e^{-3\varphi}}{4}|G_{(2)}|^2\\[1mm]
 &+\frac{\a}{2\sqrt{3}}\,\epsilon^{\mu\nu\rho\sigma}\,\left(F_{\mu\nu}F_{\rho\sigma}-\a F_{\mu\nu}G_{\rho\sigma}+\frac{\a^2}{3}G_{\mu\nu}G_{\rho\sigma}\right)\bigg]\ , 
\end{aligned}
\label{4d_action}
\end{equation}
where
\be
\begin{array}{lll}
F_{(2)} \,\equiv\, \diff \mathcal{A} +\a \,\diff \mathcal{B}\, , \hspace{1cm} & G_{(2)} \,\equiv\, \diff \mathcal{B} \, , \hspace{1cm} & \k_4^2 \,\equiv\, \dfrac{q \k_5^2}{2\p R_{S^1}} \ .
\end{array}
\ee
The 4d effective description couples the 4d metric, to two vector fields $(\mathcal{A},\mathcal{B})$ and a `complex' axio-dilaton field consisting in the two real scalars $(\varphi,\a)$. The explicit form of the 4d metric may be directly read off from the reduction ansatz \eqref{KK_Ansatz}. Since the complete metric is quite involved, we will not display it explicitly. We can provide instead its asymptotic expansion for large $r$, which is undistinguisable from the Kerr metric:
\begin{equation}
\diff s^2_4\,=\, -\left(1-\frac{2G_4 M}{r}\right)\, \diff {\hat t}^2-\frac{4G_4J\left(1-\chi^2\right)}{r}\, \diff {\hat t}\, \diff\phi+\left(1+\frac{2G_4 M}{r}\right)\,\diff r^2 +r^2 \, \diff s^2_{S^2}+\dots\,, 
\end{equation}
with mass $M$ and angular momentum $J$ given in terms of the parameters by
\begin{equation}
 M\,=\,\frac{2 r_s +r_b +\xi^t{}^2\left(r_s+2r_b\right)}{4 G_4\left(1-\xi^t{}^2\right)} \, , \hspace{1cm} J\,=\, \frac{a_s\, r_s-a_b \,r_b \, \xi^t}{2 G_4\sqrt{1-\xi^t{}^2}}\, .
\end{equation}
Note that the expression for the mass differs (apart from a normalization factor) from the five-dimensional one in \eqref{eq:5dchargesRTS}. The reason is that the 4d and 5d metrics are related by a conformal factor $e^{\varphi}$ (see \eqref{KK_Ansatz}), which contributes to the mass as a consequence of the non-trivial profile of the Kaluza-Klein scalar.\footnote{This issue does not affect the angular momentum since the conformal factor $e^{\varphi}$ between the 4d and 5d metrics does not contribute to the $1/r$ term of the ${\hat t}\phi$ component.} In addition, we further note that the mass is manifestly positive and reduces to the result of \cite{Bah:2020ogh} in the static limit, whereby $\xi^{t}\to 0$.

\begin{figure}[t]
\centering
\includegraphics[scale=0.8]{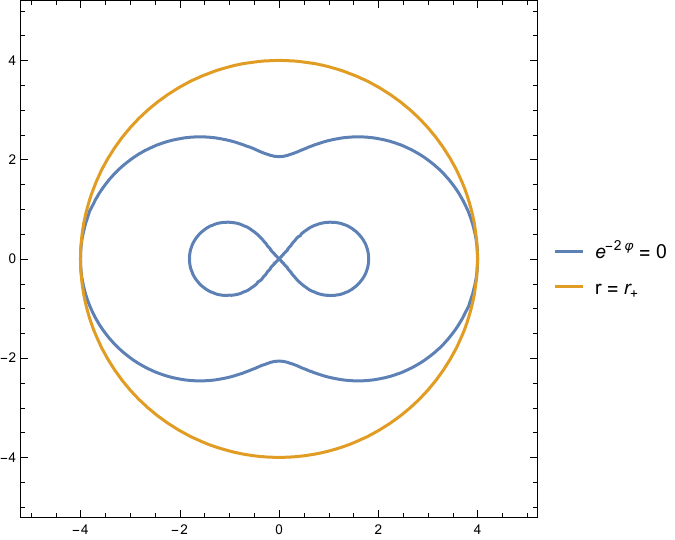}
\caption{\emph{Curve in the $\left(x, y\right)\,=\, \left(r \chi , r\sqrt{1-\chi^2}\right)$ plane along which $e^{-2\varphi}$ vanishes. We have chosen the values $r_+\,=\,4$, $r_b=2$, $r_s=1$.}}
\label{fig:singularities}
\end{figure}

Just as in the static case, the 4d geometry exhibits curvature singularities when the Kaluza-Klein scalar $\varphi$ diverges. From the explicit expression of the latter,
\be
\label{gyy}
g_{\hat{y} \hat{y}}  \,=\, e^{-2\varphi}  \,= \, 
\frac{\chi ^2 r_b r_s \left(a_b-a_s \xi ^t\right){}^2+r \Sigma  \left(r_b-r_s \xi ^t{}^2\right)+\Sigma ^2
   \left(\xi ^t{}^2-1\right)}{\Sigma ^2 \left(\xi ^t{}^2-1\right)} \,,
\ee
one sees that $\varphi$ diverges along a curve in the $(r, \chi)$ plane, which is displayed in figure~\ref{fig:singularities}. This touches the north and south poles of the $S^2$ at the cap, which are the fixed points of ${\partial}_{\hat y}$.  There, both the 4d Ricci scalar and the axionic kinetic energy diverge. 

\section{Geodetic motion}
\label{sec:geodeticmotion}

In this section we study geodetic motion of a massive spinless probe around the backgrounds discussed in the previous section. Scalar perturbations will be discussed in  section~\ref{sec:scalarperturbations}. Thanks to the three commuting isometries, one can conveniently describe the relevant dynamics in Hamilton form, 
\be
{\cal H} \,=\, g^{\mu\nu} P_\mu P_\nu = - \mu^2\, ,
\ee 
where $\mu$ is the mass of the probe and  $P_\mu = g_{\mu\nu} \dot{x}^\nu$ are the momenta. In particular,
\be 
P_r \,=\, {\Sigma \over \Delta}\,\dot{r} \,, \hspace{1cm} P_\chi = {\Sigma \over 1-\chi^2}\,\dot{\chi} \,, 
\ee 
while $E=-P_t$, $J=P_\phi$ and $p=P_y$ are conserved momenta. We shall denote them collectively by ${\cal I}_a$.
After some algebra one finds
\be
\Sigma \left({\cal H} + \mu^2\right) \,=\, \Delta P_{r}^2 +  \left(1-\chi^2\right) P_{\chi}^2 + W\left(r,\chi; {\cal I}_a\right) \,=\, 0\, ,
\ee
where  
\be
W= {N\left(r,\chi; {\cal I}_a\right) \over \Delta \left(1-\chi^2\right)} + \mu^2 \Sigma\, .
\ee
The numerator is a quartic polynomial in both $r$ and $\chi$, and is obviously quadratic in  ${\cal I}_a$ and $\mu$,
\be
 N\left(r,\chi; {\cal I}_a\right) \,=\, J^2 \Delta + B(r) \left(1-\chi^2\right) -a^2 \left(p^2-E^2\right) \Delta \left(1-\chi^2\right)^2\,,
 \ee
where $B(r)$ is again a quartic polynomial.  It is now clear that one can introduce a separation constant $K^2$, akin to Carter's constant for geodetic motion in Kerr(-Newman) black holes, as well as in circular fuzzballs \cite{Bianchi:2017sds}, and set\footnote{Note that thanks to the `negative' shift by $a^2 \left(E^2-p^2\right)$, $K^2$ is positive definite for $E\ge \sqrt{p^2+\mu^2}$ when $a^2 = a_s^2-a_b^2 \le 0$ as in the RTS regime.}
\be 
\left(1-\chi^2\right) P_\chi^2 + {J^2 \over 1-\chi^2} -a^2 \left(p^2-E^2\right)\left(1-\chi^2\right) + a^2\chi^2 \mu^2 \,=\, K^2 - a^2 \left(p^2-E^2\right)
\ee
so that
\be 
\left(1-\chi^2\right) P_\chi^2 + {J^2 \over 1-\chi^2} -a^2 \left(E^2-p^2-\mu^2\right) \chi^2 \,=\, K^2\, .
\ee
The radial Hamiltonian then becomes
\be
\Delta P_r^2 + {B(r)\over \Delta(r)} + \mu^2 r^2 \,=\, - K^2+ a^2 \left(p^2-E^2\right)\,,
\ee
which can be expressed in the very neat form
\be
P_r^2 + Q_R(r) \,=\,0\, ,
\ee
where  
\be
Q_R(r) \,=\, {N_4(r) \over \Delta(r)^2} + {\mu^2 r^2\over \Delta(r)}
\ee
and 
\be 
N_4(r) \,=\, B(r) + \left[K^2 - a^2 \left(p^2-E^2\right)\right] \Delta(r)\,.
\ee
This is again a quartic polynomial, $N_4(r)\,=\, \sum_{k=0}^4 \gamma_k r^k$, whose coefficients are given by
\be
\begin{aligned}
\gamma_4 \,=\,& p^2 - E^2 \,,\hspace{1cm} \gamma_3 \,=\, -p^2 r_s +  E^2r_b\,, \hspace{1cm}\gamma_2 \,=\, K^2-a^2\left(E^2-p^2\right) \,, \\[1mm]
\gamma_1\,=\,& a^2 \left(p^2 r_b-E^2 r_s\right)+2 J
   \left(E a_s r_s - p a_b r_b\right)-K^2 \left(r_b+r_s\right)\,, \\[1mm]
\gamma_0\,=\,&r_s r_b K^2-a^2 \left(J^2-K^2\right) -2 r_s r_b J \left(a_s  E- a_b p\right)+ \left(a_s r_b p- a_b r_s E\right)^2\, .
\end{aligned}
\ee
Geodesics are non planar in general. Planar `shear-free' geodesics at $\chi=\chi_0$ exist for specific choices of $J$ and $K$ such that $P_\chi\,=\,0$. Defining
\be
Q_{\chi}\left(\chi\right)\,=\,{J^2 \over \left(1-\chi^2\right)^2} -{ K^2 + a^2 \left(E^2-p^2-\mu^2\right) \chi^2 \over 1-\chi^2} \,, 
\ee
the `shear-free' condition boils down to
\be 
Q_{\chi}(\chi_0)\,=\, \frac{\diff Q_{\chi}(\chi)}{\diff \chi}\Big|_{\chi=\chi_0}\,=\,0\, .
\ee
It is easy to show that the equatorial plane $\chi_0=0$ is a solution of the shear-free condition if $J^2=K^2$. Assuming the angular momentum of the background is positive,  $J=+K$ corresponds to co-rotating geodesics, while the case $J=-K$ corresponds to counter-rotating geodesics.

Critical geodesics correspond to $P_r\,=\,0$ and $\frac{\diff P_r}{\diff r}\,=\,0$; they are non planar in general except for the equatorial ones.  

Let us focus for simplicity on null geodesics in the equatorial plane, which characterize the light rings \cite{Bianchi:2018kzy, Bianchi:2020des, Bacchini:2021fig, Heidmann:2022ehn}. In order to set the stage, let us recall that for static topological stars, thanks to spherical symmetry, one can focus on equatorial null geodesics. Setting for simplicity $P_y=0$, the `reduced' metric boils down to
\be
\diff s^2 = - f_s \,\diff t^2 + {\diff r^2\over f_s f_b} + r^2\,\diff \phi^2\, .
\ee
Setting $E=-P_t= -f_s \dot{t}$ and $J=P_\phi=r^2 \dot{\phi}$, the null Hamiltonian reads
\be
{\cal H} = f_sf_b P_r^2 - {E^2\over f_s} + {J^2\over r^2}\,=\,0\, .
\ee
Note that
\be
 \dot{r}= f_sf_b P_r
 \ee
vanishes for $P_r=0$ but also at $r=r_b$ ($r=r_s$ is not part of the geometry).
Critical geodesics correspond to  $Q_R\,=\,\frac{\diff Q_R}{\diff r}\,=\,0$. Introducing the impact parameter $b=\frac{J}{E}$, we have that these two conditions reduce (for $r\neq r_b$) to:
\be
Q_R = {E^2\over (r-r_s)^2(r-r_b)} [b^2 (r-r_s) - r^3]\,=\,0 
\ee
and
\be
\frac{\diff Q_R}{\diff r} = {E^2 \left(b^2 - 3 r^2\right) \over (r-r_s)^2(r-r_b)} + E^2 \left[b^2 (r-r_s) - r^3\right] \frac{\diff}{\diff r} \left[(r-r_s)^{-2}(r-r_b)^{-1}\right]\, =\, 0\,.
\ee
Using the first equation, the second one yields
\be
b_c^2\,=\,3r_c^2\,,
\ee
and plugging this into the gives 
\be
r_c^2 \left(2r_c-3r_s\right)\,=\, 0 \,.
\ee
Barring the unphysical solution $r_c=0$, one finds
\be
r_c^+= {3\over 2} r_s\,,
\ee
which may or may not be physically acceptable depending on whether $r_c$ is larger or smaller than $r_b$. If $r_b< {3\over 2} r_s$, $r_c^+= {3\over 2} r_s$ is an unstable light ring, while $r_c^-=r_b$ is a (meta)stable internal light ring. Instead, if $r_b>{3\over 2} r_s$, the only acceptable critical null geodesic has
\be
r_c^-\,=\,r_b> {3\over 2} r_s\,,
\ee
corresponding to an unstable light ring.

The situation in the rotating case should be similar, at least in the equatorial plane and provided one distinguishes between co- and counter- rotating geodesics. Indeed, setting $\chi =0$,  one has that the reduced metric is given by 
\be
\diff s^2 \,=\, - f_s \,\diff t^2 + f_b \,\diff y^2 + {r^2 \diff r^2\over \Delta} + L_\phi^2 \,\diff \phi^2 +2\Omega_t \,\diff t \,\diff\phi 
+ 2\Omega_y\, \diff y\, \diff\phi\,,
\ee
where we have introduced the quantities
\be
L_\phi^2\,=\, \frac{\left(r^2+a^2\right)^2+a^2\left(r_s r_b -\Delta\right)}{r^2}
\, ,\hspace{5mm} 
\Omega_t \,=\, -{a_s r_s \over r}
\,,\hspace{5mm}
\Omega_y \,=\, + {a_b r_b \over r}\, .
\ee
The conserved momenta now read\footnote{Note that $J=\pm K$.}
\be  
P_t\,=\,-E\,=\, - f_s \,\dot{t} + \Omega_t \,\dot{\phi }\, , \hspace{2.5mm} 
P_y\,=\,p\,=\, f_b \,\dot{y} + \Omega_y \,\dot{\phi } \, , \hspace{2.5mm} 
P_{\phi}\,=\,J\,=\, L_\phi^2 \,\dot{\phi} + \Omega_t \,\dot{t} + \Omega_y \,\dot{y}\,,
\ee
while 
\be
P_r \,=\, {r^2 \dot{r}\over \Delta }\,.
\ee
The Hamiltonian governing equatorial null geodesics can be put in the form 
 \be 
 {\cal H}\,=\, \Delta P_r^2 + W_{\rm geo}(r; E,p,J)
 \ee
 with
\be
 W_{\rm geo}(r; E,p,J) = \hat{g}^{\hat\mu\hat\nu}P_{\hat\mu} P_{\hat\nu}
 \ee
where $\hat{g}^{\hat\mu\hat\nu}$ is the reduced (inverse) metric in the $(t,y,\phi)$ isometric subspace. More explicitly, 
 one finds
 \be
 W_{\rm geo}(r; E,p,J) \,=\, E^2 {\widehat{N}_4\over \Delta} \,,
 \ee
where  $\widehat{N}_4=N_4/E^2$. Restricting to equatorial null geodesics with $p=0$, one has\footnote{Note that the condition $p=0$ can be imposed at this level notwithstanding the identifications in \eqref{eq:ident_hatted_coord}, since one is allowed to assume $p$ and $J$ to take continuous values.} 
 \be
 {\hat N}_4\,=\, r_s r_b \left(b-a_s\right)^2 + \left[r_b\left(a_s^2-b^2\right)-r_s\left(b-a_s\right)^2\right]r + \left(b^2-a^2\right)r^2 +r_b\, r^3-r^4\, ,
 \ee
where as before $b\,=\,\frac{J}{E}$.

Critical geodesics are then determined by the conditions 
 \be
 P_r\,=\, {W_{\rm geo}\over \Delta} \,=, E^2 {\widehat{N}_4\over \Delta^2}\,=\,0\,, \hspace{1cm} {\diff P_r\over \diff r} \,=\, {\diff \over \diff r}\left(E^2{\widehat{N}_4\over \Delta^2}\right)\,=\,0\,.
 \ee
 Away from the cap (where $\Delta$ vanishes), one can forget about the denominator and look simply for
 \be
 \widehat{N}_4(r_c, b_c)\,=\,0\,, \hspace{1cm}\widehat{N}_4'(r_c, b_c)\,=\,0\, .
 \ee
Eliminating quadratic terms in $b_c$ after combining the two equations, one can express $b_c$ in terms of $r_c$ and then get a quintic equation for $r_c$ whose solutions cannot in general be found by radicals.

For small $a$ and thus\footnote{Recall that $a^2=a_s^2-a_b^2$ is negative in the RTS regime.} $a_s= \sqrt{-a^2} {r_s\over r_b-r_s}$ and $a_b= \sqrt{-a^2} {r_b\over r_b-r_s}$ one can expand the solution around the non-rotating case. To first order in $a$, $b_c^2 = 3 r_c^2$ still holds and one finds
\be 
r_c^{\pm} = {3r_s\over 2} \pm {2a_s\over \sqrt{3}} + \dots
\ee
with the $+$ sign for counter-rotating geodesics and $-$ sign for co-rotating geodesics. Moreover one also has 
\be
\tilde{r}_c^{\pm} = \pm {2a_s\over \sqrt{3}}\, .
\ee
Barring the negative solution, the positive solution is acceptable if $2a_s>\sqrt{3}r_+$ which is beyond the present approximation scheme and requires a more detailed (numerical) analysis, along the lines for instance of \cite{Cipriani:2025ini}.

Finally one can check that at $r=r_+$  (where $\Delta$ vanishes) due the potential barrier one cannot find any critical geodesics.
 
%

\section{Scalar perturbations}
\label{sec:scalarperturbations}
Static topological stars are linearly stable under scalar and metric perturbations \cite{Heidmann:2023ojf,Bianchi:2023sfs,Bianchi:2024vmi,Bianchi:2024rod,Cipriani:2024ygw,Dima:2024eqq,Bena:2024hoh,DiRusso:2025lip,Dima:2025zot}. In both cases, the angular and radial parts of the perturbations can be indeed separated into two decoupled ordinary differential equations. Being spherically-symmetric solutions, the angular equation can be solved in terms of standard spin-weighted harmonics. The radial equation can be put in the form of a confluent Heun and generalized confluent Heun type (for scalar and gravitational perturbations respectively). As it turns out, all QNMs frequencies have negative imaginary part. This is in contrast with what one finds for other horizonless smooth geometries like JMaRT \cite{Cardoso:2005gj, Bianchi:2023rlt}.

 In this section, we study scalar perturbations of the rotating black string and topological star backgrounds, showing that the equation describing the dynamics of such perturbations separates into two ordinary differential equations for the radial and angular parts of the perturbations, which are instances of confluent Heun type. For simplicity, we focus on the massless case:
\begin{equation}
\label{eq:wave_eq}
\Box \,  \Phi(x)\,=\,{1\over \sqrt{-g}} \partial_\mu\left(\sqrt{-g} g^{\mu\nu} \partial_\nu\right)\Phi(x) \,=\, 0\, .
\end{equation}
Exploiting the three commuting isometries, one can take as an ansatz
\be
\label{eq:ansatz_Phi}
\Phi(x)\,=\,e^{-{\rm i} {\omega} {t}+{\rm i} m \phi+{\rm i} {p} { y}}S\left(\chi\right) R\left(r\right)\,,
\ee
where $m\in {\mathbb Z}$. As we showed in section~\ref{sec:RTS}, the soliton only exists (when completely smooth) if the coordinates satisfy twisted periodic identifications \eqref{ident3}, or equivalently  \eqref{eq:ident_hatted_coord}. This implies the following quantization must hold
\begin{equation}
\label{eq:quantization_p}
\left(p-\xi^t\omega\right) R_y-\frac{m h}{q} \in {\mathbb Z}\, .
\end{equation}
In the black string case, however, we do not need to impose twisted boundary conditions, which means that the quantization condition for $p$ simply reads (assuming $y\sim y+2\pi R_y$)
\begin{equation}
pR_y\,\in\, {\mathbb Z}\,.
\end{equation}

Plugging now \eqref{eq:ansatz_Phi} into \eqref{eq:wave_eq}, one finds that the wave equation separates into the following two ordinary differential equations:
\bea
{\diff\over \diff\chi} \left[\left(1-\chi ^2\right) {\diff S\left(\chi\right)\over \diff\chi} \right]  +\left[A-\frac{m^2}{1-\chi
   ^2} + a^2 \chi ^2 \left({\omega}^2-{p}^2\right)\right] S\left(\chi\right) &\,=\,&0\, , \\
{\diff\over \diff r} \left[ \Delta(r)  {\diff R(r)\over \diff r}\right]
 {+} \left[ {-}A{+}\frac{ {\cal C}(r) }{\Delta(r) }\right]R(r) &\,=\, & 0  \, ,  \label{heun} 
 \eea
being $A$ a (Carter-like) separation constant and ${\cal C}(r)=\sum_{k=0}^4 \lambda_k r^k$ a fourth-order polynomial with coefficients
 \begin{equation}
 \begin{aligned}
\lambda _0 \,=\,&  - r_b r_s \left(p^2 a_b+\omega ^2 a_s\right) +m^2a^2+2
   r_b r_s \left(-m p a_b+p \omega  a_b a_s+m \omega  a_s\right) \,,\\[1mm]
\lambda _1\,=\,&  a^2 \left(\omega ^2  r_s-p^2 r_b\right)+2 m \left(p a_b r_b- \omega  a_s r_s\right)\,,\\[1mm] 
\lambda _2 \,=\,& a^2 \left(\omega^2-p^2\right) \,,\\[1mm]
\lambda _3\,=\,& p^2 r_s-\omega ^2 r_b \,, \\[1mm] 
\lambda _4\,=\,& {\omega}^2-{p}^2\, .
\end{aligned}
\end{equation}
Note that up to a sign and the shift by terms in $A$, the quartic polynomial ${\cal C}(r)$ coincides with $N_4(r)$, appearing in the radial effective potential for null geodesics, after replacing $A$ with $K^2$, $m$ with $J$ and $\omega$ with $E$.  
Both equations in (\ref{heun}) are confluent Heun equations. The angular equation exhibits regular singularities at $\chi=\pm 1$ and an irregular singularity at infinity. 
Observe that this equation coincides with the one describing the polar angular dynamics in Kerr for $p\,=\, 0$, with the important difference  that now $a^2$ can be either positive or negative, depending on whether we are inthe black string or soliton regime, respectively.
 
 On the other hand, the radial equation has regular singularities at $r=r_\pm$ and an irregular one at infinity.  Following the recently proposed gauge(CFT)/gravity correspondence for black holes, fuzzballs, and (lo and behold) cosmological perturbations \cite{Aminov:2020yma,Bianchi:2021xpr,Bianchi:2021mft,Bonelli:2021uvf,
Bonelli:2022ten,Bianchi:2022qph,Consoli:2022eey,Bautista:2023kns,Bianchi:2024zgn,DiRusso:2024rgi,Cipriani:2025ikx}, one can relate the Heun equation and its confluences to the quantum Seiberg-Witten (qSW) curves describing a $SU(2)$ gauge theory with up to four fundamental hyper-multiplets. 

For three hypers, the qSW curve can be written as 
  \be
  \label{quantumsw}
  \left[ z^2 P_L(-z\partial_z+\ft12) - z  P(-z\partial_z) + x \,  P_R(-z\partial_z-\ft12) \right] {\cal Z}(z)\,=\,0\,, 
  \ee
 with 
 \bea
 P_L(v)&=& (v-m_1)(v-m_2) \,, \hspace{1cm}  P_R(v)\,=\,v-m_3\,, \nn\\[1mm]
  P(v) &=& v^2-u+x(v+\ft12-m_1-m_2-m_3)\,.
 \eea
Performing a change of variable, 
 \be
  {\cal Z}(z)={e^{x\over 2z} \Psi(z) \over \sqrt{ z (1-z)^{1+m_1+m_2}}} \, ,
  \ee
we can bring \eqref{quantumsw} to Schr\"odinger-like canonical form
\be
\label{canonical}
\Psi(z) + Q(z)\Psi(z)\,=\,0\,,   
\ee
where
 \be
   Q(z) \,=\, {-}\frac{x^2}{4 z^4}{+}\frac{x m_3}{z^3}{+}\frac{1{-} \left(m_1{-}m_2\right)^2}{4(z{-}1)z}+\frac{ 1{-}
   \left(m_1{+}m_2\right){}^2}{4(z-1)^2 z}  +\frac{u{-}\ft{1}{4}{+}\ft{x}{2}
   \left(m_1{+}m_2{-}1\right)}{(z{-}1)
   z^2} \, .
 \ee
The angular and radial equations in \eqref{heun} can be also put into the  Schr\"odinger-like canonical form \eqref{canonical}. To this aim, one has to identify 
\begin{equation}
\begin{aligned}
\chi \,=\,& {2\over z}-1 \,, \hspace{5mm} S(\chi)\,=\, {\Psi(z) \over \sqrt{2-2z} }\, , \hspace{5mm}  x\,=\,4 a \sqrt{\omega^2-p^2}  \,, \\[1mm]
 m_1 \,=\,&  m \,, \hspace{5mm}  m_{2,3}\,=\, 0 \,, \hspace{5mm} u\,=\, A{+}\ft{1}{4}{+}a  (a{-}2 m{+}2)\sqrt{\omega^2{-}p^2} \,, 
\end{aligned}
\end{equation}
in the angular case, and
\begin{equation}
\begin{aligned}
r \,=\,& r_- + {r_+ - r_-\over z} \,, \hspace{5mm} R(r)\,=\,{\Psi (z) \over \sqrt{  2-2z } }\,, \hspace{5mm}  x\,=\, 2 \ii \sqrt{\lambda _4} \left(r_--r_+\right) \, , \\[1mm]
 m_{1,2} \,=\,&  \frac{2 \ii}{r_+-r_-}\left(\sqrt{{\cal\ C}(r_+)}\mp \sqrt{{\cal\ C}\left(r_-\right)}\right) \,, \hspace{5mm} m_3\,=\,  \frac{\ii}{2 \sqrt{\lambda _4}} \left[\lambda _3+2 \lambda _4
   \left(r_-+r_+\right)\right]\,,   \\[1mm]
u \,=\,&  A+\ft{1}{4}-\lambda _2-\lambda _4 \left(r_-^2-2  r_+ r_- +3  r_+^2\right) +i \sqrt{\lambda _4} \left(r_- - r_+\right)  -\lambda _3 \left(r_-+2
   r_+\right) \\[1mm]
&-2 \sqrt{\lambda _4\,  {\cal C}(r_+)}\, ,
\end{aligned}
\end{equation}
 in the radial case.
 
 \subsection{$r_s=r_b$ case}
 
A significant simplification of the equations describing linear perturbations takes place in the extremal case $r_s=r_b$. Along this line, $a^2=0$ and the angular equation can be explicitly solved in terms of spherical harmonics with separation constant,
  \be
  A=\ell(\ell+1)  \, .
  \ee
For $\mu=0$, the radial equation becomes:
   \be
 {\diff\over \diff r} \left[\left(r{-}r_s\right){}^2 \frac{\diff R}{\diff r}\right] {+}R(r) \left[ \frac{2 m a_s \left(p-\omega\right) r_s}{r-r_s}{+}\frac{\left(\omega ^2-p^2\right) \left(a_s^2 r_s^2{+}r^4{-}r^3
   r_s\right){+}2 p \omega  a_s^2 r_s^2}{\left(r{-}r_s\right){}^2} -A\right]=0\, .
 \ee
This corresponds to a doubly confluent Heun equation with irregular singularities at $r=r_s,\infty$. It can be related to the
 quantum SW curve of a ${\cal N}=2$ $SU(2)$ gauge theory with two fundamental hyper-multiplets. The SW curve is given again by \eqref{quantumsw}, with
\begin{equation}
P_L(v)\,=\, v-m_1 \, ,\hspace{5mm} P_R(v)\,=\,v-m_2 \, ,\hspace{5mm} P(v)\,=\, v^2-u+x \,.
 \end{equation}
 It can be brought to canonical form \eqref{canonical} via the change of variable
 \be
 {\cal Z}= \sqrt{z} \, e^{ {x\over 2z}-{z\over 2}   } \, \Psi(z) 
 \ee
where now $Q(z)$ is given by
 \be
 Q(z) \,=\,-\frac{x^2}{4 z^4}+\frac{m_2 x}{z^3}+\frac{m_1}{z}+\frac{2 x -4 u+1}{4 z^2}-\frac{1}{4}\, .
 \ee
 The `gauge-gravity dictionary' for the radial equation becomes
 \bea
 r&=& r_s+ {1 \over \nu z}  \qquad, \qquad  x= a_s r_s(p-\omega ) \sqrt{p^2-\omega^2 }  \quad , \quad 
 \nu = \frac{1}{2  a_s r_s(p-
   \omega)}  \nn\\
 m_1&=&  m-\frac{(p+\omega ) r_s^2}{2 a_s} \qquad , \qquad m_2= -\frac{3}{2} r_s\sqrt{p^2-\omega^2 } 
    \nn\\
   u&=&  A+\ft12  +2 a_sr_s (p-\omega ) \sqrt{p^2-\omega^2 } +3 (p^2-\omega^2) r_s^2 
 \eea
For $p\,=\, 0$ and $m\,=\,0$, the radial equation can be written in the Schr\"odinger-like canonical form
  \be
 \Psi''(r)+Q_r  \Psi(r)=0 \,, \hspace{1cm} R(r)\,=\,{\Psi(r)\over r-r_s} 
 \ee
 with
\be
Q_r\,=\,\frac{\omega ^2 \left(a_s^2 r_s^2{+}r^4{-}r^3
   r_s\right)}{\left(r{-}r_s\right){}^4} -\frac{\ell\left(\ell+1\right)}{\left(r-r_s\right)^2}\, .
\ee 
The simple form of $Q_r(r)$ allows us to explicitly determine the location $r_c$ and the critical frequency $\omega_c$ associated the light-rings of the solution. 
They are given by solutions of the critical equations
\be
Q_r(r_c, \omega_c)= \partial_r Q_r(r_c, \omega_c)\,=\,0\, .
\ee
  In the limit of $a_s$ small one finds
\be
 r_c = \frac{3 r_s}{2}-\frac{8 a_s^2}{9 r_s} +\ldots \,, \hspace{1cm} \omega_c =\frac{2 \sqrt{\ell (\ell+1)} }{3 \sqrt{3} r_s}  \left(1+  \frac{8   a_s^2}{27 r_s^2}+\ldots \right) \,,
 \ee
 with dots denoting higher-order terms in $a_s$. 
 
\section*{Acknowledgments}

We thank Mohammad Akhond, Iosif Bena, Donato Bini, Giulio Bonelli, Giorgio Di Russo, Francesco Fucito, Stefano Giusto, Paolo Pani, Nicol\`o Petri, Rodolfo Russo, Raffaele Savelli and Enrico Turetta for stimulating discussions, and to Pierre Heidmann for useful correspondence. We acknowledge support by INFN through the network ST\&FI ``String Theory \& Fundamental Interactions'' and by the MIUR-PRIN contract 2020KR4KN2 {\it ''String Theory as a bridge between Gauge Theories and Quantum Gravity''}, within which A.~R. holds a postdoctoral fellowship.

\bibliography{references.bib}
\bibliographystyle{JHEP}
\end{document}